\documentclass[11pt,draftcls,onecolumn, technote]{IEEEtran}
\usepackage[dvips]{epsfig}
\usepackage[dvips]{graphicx}
\usepackage{boxedminipage}
\usepackage{color}
\usepackage{amsfonts}
\usepackage{amssymb}
\usepackage{amsmath}
\usepackage{graphics,amsmath,amsfonts,amssymb,epsfig,latexsym,amsfonts,graphicx,amssymb}
\usepackage{eqparbox}
\usepackage{cite}
\usepackage{setspace}

\newcommand{\bfbeta}{{\mbox{\boldmath $\beta$}}}

\newcommand{\bfPhi}{{\mbox{\boldmath $\Phi$}}}

\newcommand{\bx}{\mathbf{x}}
\newcommand{\bs}{\mathbf{s}}

\newcommand{\bR}{\mathbf{R}}

\newcommand{\bI}{\mathbf{I}}

\newcommand{\bc}{\mathbf{c}}

\newcommand{\ba}{\mathbf{a}}
\newcommand{\bp}{\mathbf{p}}

\newcommand{\be}{\mathbf{e}}

\newcommand{\cB}{\mathcal{B}}
\newcommand{\cP}{\mathcal{P}}
\newcommand{\cE}{\mathcal{E}}

\newcommand{\cK}{\mathcal{K}}
\newcommand{\cW}{\mathcal{W}}
\newcommand{\cQ}{\mathcal{Q}}

\newcommand{\cN}{\mathcal{N}}

\newcommand{\ctN}{\widetilde{\mathcal{N}}}
\newcommand{\bhI}{\widehat{\mathbf{I}}}

\newcommand{\bhp}{\widehat{\mathbf{p}}}
\newcommand{\btp}{\widetilde{\mathbf{p}}}
\newcommand{\bha}{\widehat{\mathbf{a}}}
\newcommand{\bta}{\widetilde{\mathbf{a}}}

\newcommand{\tp}{\widetilde{{p}}}

\def\BibTeX{{\rm B\kern-.05em{\sc i\kern-.025em b}\kern-.08em
    T\kern-.1667em\lower.7ex\hbox{E}\kern-.125emX}}

\begin{document}
\title{Joint Access Point
Selection and Power Allocation for Uplink Wireless Networks}
\author{Mingyi Hong, Alfredo Garcia, Jorge Barrera and Stephen G. Wilson
\vspace{-0.8cm}
\thanks{M. Hong, A. Garcia and J. Barrera are with the
Department of Systems and Information Engineering, University of
Virginia. S. G. Wilson is with the Department of Electrical and
Computer Engineering, University of Virginia.}
\thanks{Part of this paper was presented in
IEEE INFOCOM 2011 \cite{hong11_infocom}.}}

\maketitle
\begin{abstract}
We consider the distributed uplink resource allocation problem in a
multi-carrier wireless network with multiple access points (APs).
Each mobile user can optimize its own transmission rate by selecting
a suitable AP and by controlling its transmit power. Our objective
is to devise suitable algorithms by which mobile users can {\em
jointly} perform these tasks in a distributed manner. Our approach
relies on a game theoretic formulation of the joint power control
and AP selection problem. In the proposed game, each user is a
player with an associated strategy containing a discrete variable
(the AP selection decision) and a continuous vector (the power
allocation among multiple channels). We provide characterizations of
the Nash Equilibrium of the proposed game, and present a set of
novel algorithms that allow the users to efficiently optimize their
rates. Finally, we study the properties of the proposed algorithms
as well as their performance via extensive simulations.
\end{abstract}

\vspace{-0.4cm}
\section{Introduction}
In this paper we consider the {joint} access point (AP) selection
and power allocation problem in a wireless network with multiple APs
and mobile users (MUs). The APs operate on non-overlapping spectrum
bands, each of which is divided into parallel channels. Each MU's
objective is to find a suitable AP selection, followed by a
vectorial power optimization.

The problem of joint AP selection and power control belongs to the
category of cross-layer resource allocation in wireless
communications. Such problem is important in many practical
networks. For example, in the IEEE 802.22 cognitive radio Wireless
Regional Area Network (WRAN) \cite{stevenson09}, a geographical
region may be served by multiple service providers (SPs), or by
multiple APs installed by a single SP \cite{acharya09}. The users
transmit by sharing the spectrum offered by a SP/AP. Moreover the
users enjoy the flexibility of being able to dynamically select the
less congested SP/AP. Another example in which such joint
optimization plays an important role is the heterogenous networks
\cite{Niyato05}, \cite{McNair04}, in which different technologies
such as Wi-Fi, 3G, LTE or WiMAX are available for the same region.
The MUs can choose from one of these technologies for communication,
and they can switch between different technologies to avoid
congestion (the so called ``vertical handoff", see \cite{McNair04}).
As suggested in \cite{apcan06} and \cite{saraydar01}, compared with
the traditional closest AP assignment strategy, it is generally
beneficial, in terms of the system-wide performance and the
individual utilities, to include the AP association as the users'
decision variable whenever possible.

Distributed strategies for resource allocation in multi-carrier or
single-carrier network with a single AP have been extensively
studied. In \cite{lai08}, the authors consider the uplink power
control problem in a fading multiple access channel (MAC). A
water-filling (WF) game is devised in which the users compete with
each other to maximize their individual rates. The authors show that
the MAC channel capacity is achieved when all the users have {\it
perfect} knowledge of all other users' channel state information
(CSI). Such assumption, however, is unrealistic in a distributed
setting. Reference \cite{he08} proposes a distributed power
allocation scheme for uplink OFDM systems where the channel state is
modeled to having only discrete levels. Notably, this scheme only
requires that each user has the knowledge of {\it its own} CSI,
which greatly reduces the signaling overhead. Reference \cite{yu04}
considers a generalization of a multi-carrier MAC channel, and
proposes a distributed iterative water-filling (IWF) algorithm to
compute the maximum sum capacity of the system. 

The problem of joint AP selection and power control has been
addressed in single-carrier cellular networks as well. References
\cite{hanly95} and \cite{yates95b} are early works that consider
this problem in an uplink spread spectrum cellular network. The
objectives are to find an AP selection and power allocation tuple
that minimizes the total transmit powers while maintaining the
users' quality of service requirements. The authors of
\cite{apcan06} and \cite{saraydar01} cast a similar problem (with an
objective to minimize the individual cost) into a game theoretical
framework, and propose algorithms to find the Nash Equilibrium (NE)
of their respective games. All of these papers consider the {\it
scalar} power allocation problem. The work presented in this paper
differs significantly in that a more difficult {\it vector} power
allocation problem is considered, in which the users can potentially
use all the channels that belong to a particular AP.

There are a few recent works addressing related joint power control
and band selection or channel selection problems in multicarrier
networks. Reference \cite{Meshkati06} proposes to use game theory to
address the problem of distributed energy-efficient channel
selection and power control in an uplink multi-carrier CDMA system.
Again the final solution mandates that the users choose {\it a
single} optimum channel as well as {\it a scalar} power level to
transmit on the selected channel. Reference \cite{acharya09}
considers the uplink vector channel power allocation and spectrum
band selection problem in a cognitive network with multiple service
providers. The users can select the size of the spectrum as well as
the amount of power for transmission. However, the authors avoid the
difficult combinatorial aspect of the problem by assuming that the
users are able to connect to multiple service providers at the same
time. Such assumption may induce considerable signaling overhead on
the network side as well as hardware implementation complexity on
the user device \footnote{In WLAN literature, such network is also
referred to as ``multi-homing" network, see \cite{shakkottai07} and
the reference therein.}.

To the best of our knowledge, this is the first work that proposes
distributed algorithms to deal with joint AP selection and power
allocation problem in a multi-channel multi-AP network. We first
consider the single AP configuration, and formulate the uplink power
control problem into a non-cooperative game, in which the MUs
attempt to maximize their transmission rates.  An iterative
algorithm is then proposed that enables the MUs to reach the
equilibrium solutions in a distributed fashion. We then incorporate
the AP selection aspect of the problem into the game. In this more
general case, each user can select not only its power allocation,
but its AP association as well. Although non-cooperative game theory
has recently been extensively applied to solve resource allocation
problems in wireless networks (e.g.,
\cite{leshem09,game09,hong12survey}), the considered game is
significantly more complex due to its mixed-integer nature. We
analyze the NE of the game, and develop a suite of algorithms that
enable the MUs to identify equilibrium solutions in a distributed
fashion.

This paper is organized as follows. In Section
\ref{secProblemFormulation}, we consider the power allocation
problem in a single AP network. From Section \ref{secSystemJoint} to
Section \ref{secJJASPA}, we discuss the problem of joint AP
selection and power allocation. We introduce a game theoretic
formulation of the problem and propose algorithms to reach the NE of
the game. In Section \ref{secSimulation}, we show the numerical
results. The paper concludes in Section \ref{secConclusion}.

{\it Notations}: We use bold faced characters to denote vectors. We
use $\bx[i]$ to denote the $i$th element of vector $\bx$. We use
${\mathbf{x}_{-i}}$ to denote the vector
$\left[\mathbf{x}[1],\cdots,\mathbf{x}[i-1],\mathbf{x}[i+1],\cdots
\mathbf{x}[N]\right]$. We use $\mathbf{e}_{j}$ to denote a unit
vector with all entries $0$ except for the $j$th entry, which takes
the value $1$; use $\mathbf{e}$ to denote the all $1$ vector. We use
$[y,\mathbf{x}_{-i}]$ to denote a vector $\mathbf{x}$ with its $i$th
element replaced by $y$.  The key notations used in this paper are
listed in Table \ref{tableSymbols}.

\begin{table*}[htb]
\begin{center}
{\small \vspace*{-0.5cm}
\begin{tabular}{|c |c | c|c|}
\hline
 $I^k_i$ & Total
 interference for MU $i$ on channel $k$ & $\bI_{i,w}$ &The collection $\{I^k_i\}_{k\in\mathcal{K}_w}$
\\
\hline $\mathbf{a}$& Association
profile& $\mathbf{a}_{-i}$ & Association profile without MU $i$\\
 \hline
$|h^k_{i,w}|^2$& Channel gain for MU $i$ on channel $k$ with AP
$w$ &$\bfPhi_i(\bI_i;w)$ & Vector WF operator for MU $i$\\
\hline $p^k_{i,w}$ & Power for MU $i$ on channel $k$ with
AP $w$ &$\bp_{i,w}$& The collection $\{p^k_{i,w}\}_{k\in\cK_w}$\\
\hline
$\bp^*(\ba)$& System NE power allocation for given $\ba$ &$\bp_w$& The collection $\{\bp_{i,w}\}_{i:\ba[i]=w}$\\
 \hline
 $P_w(\bp_w;\ba)$& Potential function for AP $w$&$P(\bp;\ba)$&System Potential Function\\
 \hline
 $\cE(\ba)$&The set of solutions for maximizing $P(\bp;\ba)$&$\bar{P}(\ba)$& Maximum value of $P(\bp;\ba)$\\
 \hline
 $R_i(\bp_{i,w},\bp_{-i,w};\ba)$& User $i$'s rate when connected with AP $w$&$\cN_w$& The set of all users with AP $w$\\
\hline
\end{tabular} } \label{tableSymbols}
\caption{\small The Summary of Notations }\vspace*{-0.7cm}
\end{center}
\vspace*{-0.5cm}
\end{table*}

\vspace{-0.3cm}
\section{Optimal Power Allocation in a Single AP Network}\label{secProblemFormulation}
In this section, we briefly consider the simpler network
configuration with a single AP. We provide insights into the
relationship between optimal and equilibrium power control
strategies of the users. We study algorithms whose convergence
properties will be useful in our subsequent study of the multiple AP
configuration.

\vspace{-0.3cm}
\subsection{System Model}
Consider a network with a single AP. The MUs are indexed by a set
$\mathcal{N}\triangleq\{1,2,\cdots,N\}$. Normalize the available
bandwidth to $1$, and divide it into $K$ channels. Define the set of
available channels as $\mathcal{K}\triangleq\{1,2,\cdots,K \}$.

Let $x^k_i$ denote the signal transmitted by MU $i$ on channel $k$;
let $p^k_i=\mathbb{E}\big[|x^k_i|^2\big]$ denote the transmit power
of MU $i$ on channel $k$. Let $z^k\sim CN(0, n^k)$ denote the white
complex Gaussian thermal noise experienced at the receiver of AP
with mean zero and variance $n^k$. Let $h_i^k$ denote the channel
coefficient between MU $i$ and the AP on channel $k$. The signal
received at the AP on channel $k$, denoted by $y^k$, can then be
expressed as: $ y^k=\sum_{i=1}^{N}x_i^k h_i^k+z^k$. Let
$\mathbf{p}_i\triangleq\left[p^1_i,\cdots,p^K_i\right]^{\intercal}$
be MU $i$'s transmit power profile; let
$\mathbf{p}\triangleq\left[\mathbf{p}^{\intercal}_1,\cdots,\mathbf{p}^{\intercal}_{N}\right]^{\intercal}$
be the system power profile. Define ${P}_i$ to be MU $i$'s maximum
allowable transmit power. The set of feasible transmit power vectors
for MU $i$ is
$\mathcal{P}_i\triangleq\left\{\mathbf{p}_i:\mathbf{p}_i\ge\mathbf{0},
\sum_{k=1}^{K}p_i^k\le {P}_i\right\}$.

Assume that the AP is equipped with single-user receivers, which
treat other MUs' signals as noise when decoding a specific MU's
message.  This assumption allows for implementation of
low-complexity receivers at the AP, and it is generally accepted in
designing distributed algorithm in the MAC, see
\cite{lai08,hong12survey}. Under this assumption, the MU $i$'s
achievable rate can be expressed as \cite{cover05}:{\small
\begin{align}
R_i(\mathbf{p}_i,\mathbf{p}_{-i})=\frac{1}{K}\sum_{k=1}^{K}\log\left(1+\frac{{p}_i^k|h_i^k|^2}{n^k+\sum_{j\ne
i}{p}_j^k|h_j^k|^2}\right).\label{eqRate}
\end{align}}
Clearly the rate $R_i(\cdot)$ given above represents the information
theoretical rate that can be achieved only when using Gaussian
signaling. This rate serves as the performance upper bound for any
practically achievable rates. It is widely used for the design and
analysis of resource allocation and spectrum management schemes in
wireless networks, see e.g., \cite{lai08,luo08a,hong12survey}.

Another important assumption is that each MU $i\in\mathcal{N}$ knows
its own channel coefficients $\{h_i^k\}_{k\in\mathcal{K}}$ and the
sum of noise plus interference $\sum_{j\ne i}p_j^k|h_j^k|^2+n^k$ on
each channel $k$. These pieces of information can be fed back by the
APs \cite{Meshkati06}.

Assume that time is slotted, and there is network-wide slot
synchronization. Such synchronization can be made possible by
equipping the users with GPS devices. The MUs can adjust their power
allocation in a slot by slot basis. The task for each MU is then to
compute its optimal power allocation policy in a distributed manner.
In the following subsection, we formulate such problem into a
game-theoretical framework.

\vspace{-0.3cm}
\subsection{A Non-Cooperative Game
Formulation}\label{subSingleAPGame}

To facilitate the development of a distributed algorithm, we model
each MU as a selfish agent, and its objective is to maximize its own
rate. More specifically, MU $i$ is interested in solving the
following optimization problem:{\small
\begin{align}
&\max_{\mathbf{p}_i\in\mathcal{P}_i}~\frac{1}{K}\sum_{k=1}^{K}\log\left(1+\frac{{p}_i^k|h_i^k|^2}
{n^k+I^k_i}
\right)
\label{eqUserProblem}
\end{align}}
where $I^k_i\triangleq\sum_{j\ne i}{p}_j^k|h_j^k|^2$ is the total
interference MU $i$ experiences on channel $k$. The solution to this
problem, denoted as $\bp^*_i$ is the well-known single-user WF
solution, which is a function of $\{I^k_i\}_{k=1}^{K}$
\cite{cover05}:{\small
\begin{align}
(p_i^k)^*=
\left[\sigma_i-\frac{n^k+I^k_i}{|h_i^k|^2}\right]^{+}\triangleq\Phi^k_i(I^k_i),~\forall~k\in\mathcal{K}
\label{eqWaterFilling}
\end{align}}
\hspace{-0.1cm}where $\sigma_i\ge 0$ is the dual variable for the
power constraint. Let $\bI_i\triangleq\{I_i^k\}_{k=1}^{K}$ be the
interference experienced by MU $i$ on all channels. We can define
the vector WF operator $\bfPhi_i(\bI_{i})$ as:{\small
\begin{align}
\bfPhi_i(\bI_{i})\triangleq[\Phi^1_i(I^1_{i}),\cdots,\Phi^K_i(I^K_{i})].\label{eqWaterFillingVector}
\end{align}}
We introduce a non-cooperative power control game where {\it i)} the
players are the MUs; {\it ii)} the utility of each player is its
achievable rate; {\it iii)} the strategy of each player is its power
profile. We denote this game as
$G=\{\mathcal{N},\mathcal{P},\{R_i(\cdot)\}_{i\in\mathcal{N}}\}$,
where $\mathcal{P}=\prod_{i\in\mathcal{N}}\mathcal{P}_i$ is the
joint feasible region of all MUs. The NE of game $G$ is the
strategies $\{\mathbf{p}_i^*\}_{i\in\mathcal{N}}$ satisfying
\cite{osborne94}:{\small
\begin{align}
\mathbf{p}^{*}_i\in \arg\max_{\mathbf{p}_i\in\mathcal{P}_i}
R_i(\mathbf{p}_i,\mathbf{p}^*_{-i})~\forall~i\in\mathcal{N}.\label{eqDefineNE}
\end{align}}
Intuitively, a NE of the game is a stable point of the system in
which no player has the incentive to deviate from its strategy,
given the strategies of all other players. To analyze the NE, let us
introduce the {\em potential function} $P: \mathcal{P}\to
\mathbb{R}$ as follows:{\small
\begin{align}
P(\mathbf{p})\triangleq
\frac{1}{K}\sum_{k=1}^{K}\left(\log\left(n^k+\sum_{i=1}^{N}|h_i^k|^2p_i^k\right)-\log(n^k)\right).\label{eqPotential}
\end{align}}
We can readily observe that for any $\mathbf{p}_i$ and
$\bar{\mathbf{p}}_i\in\mathcal{P}_i$ and for fixed
$\mathbf{p}_{-i}$, the following identity is true{\small
\begin{align}
R_i(\mathbf{p}_i,\mathbf{p}_{-i})-R_i(\bar{\mathbf{p}}_i,\mathbf{p}_{-i})
=P(\mathbf{p}_i,\mathbf{p}_{-i})-P(\bar{\mathbf{p}}_i,\mathbf{p}_{-i})\label{eqPotentialProperty2}.
\end{align}}
Due to the property \eqref{eqPotentialProperty2}, the game $G$ is
referred to as a {\it cardinal potential} game.  Note further that
the potential function $P(\bp)$ is concave in $\bp$. The following
theorem is a classical result in game theory (see, e.g.,
\cite{monderer96}).

\newtheorem{T1}{Theorem}
\begin{T1}\label{theoremPotential}
{\it A potential game with concave potential and compact action spaces admits at
least one pure-strategy NE. Moreover, a feasible
strategy is a NE of the game if and only if it
maximizes the potential function.}
\end{T1}

In light of the above theorem, we immediately have the following
corollary.
\newtheorem{C1}{Corollary}
\begin{C1}\label{corPotential}
{\it $\mathbf{p}^*$ is a NE of the game $G$ if and only if{\small
\begin{align}
\hspace{-0.1cm}\mathbf{p}^*\in\arg\max_{\mathbf{p}\in\mathcal{P}}
\frac{1}{K}\sum_{k=1}^{K}\left(\log\left(n^k+\sum_{i=1}^{N}|h_i^k|^2p_i^k\right)-\log(n^k)\right).
\label{eqOptimizationPotential}
\end{align}}}
\end{C1}

We note that the potential game formulation for a single AP network
is not entirely new. It can be easily generalized from the existing
results such as \cite{scutari06}. The main purpose of going through
this formulation in detail is to facilitate our new formulation for
the multiple AP setting in later sections.

Interestingly, the maximum value of the potential function
\eqref{eqPotential} equals to the maximum achievable sum rate of
this $N$-user $K$-channel network. Such equivalence can be derived
by comparing the expression for the sum capacity of this
multichannel MAC with the potential \eqref{eqPotential}. This
observation indicates that at a NE of the game $G$, the MUs'
transmission strategy is capacity achieving. However, in general the
NE is still {\it inefficient}, meaning that the sum of the
individual MUs' rates is less than the maximum sum rate of the
system, as only suboptimal single-user receivers are used at the AP
\footnote{ In general, one needs to have both {\it optimal
transmission and optimal receiving strategies} to achieve the MAC
capacity, where the optimal receiving strategy is to perform the
successive interference cancelation, see \cite{cover05,yu04}. }.

One exception to such inefficiency is when the NE represents an FDMA
strategy, in which there is at most one MU transmitting on each
channel. Obviously single-user receiver is optimal in this case, as
there is no multiuser interference on any channel. Reference
\cite{Ohno02multicarriermultiple} shows that when the number of
channels tends to infinity, {\it any} optimal transmission strategy
is FDMA for a multichannel MAC channel. In our context, this result
says when $K$ becomes large, any NE of the game $G$ represents an
efficient FDMA solution.

\vspace{-0.3cm}
\subsection{The Proposed Algorithms and
Convergence}\label{subsecPowerAlgorithm}

We have established that finding the NE of the game $G$ is
equivalent to finding
$\mathbf{p}^*\in\arg\max_{\mathbf{p}\in\mathcal{P}} P(\mathbf{p})$,
which is a convex problem and can be solved in a centralized way.
However, when the MUs are selfish and uncoordinated, it is not clear
how to find such NE point in a distributed fashion.

Reference \cite{yu04} proposes a distributed algorithm named
sequential iterative water filling (S-IWF) to compute the solution
of the problem $\max_{\mathbf{p}\in\mathcal{P}} P(\mathbf{p})$. In
iteration $t$ of the S-IWF, a {\it single MU} $i$ updates its power
by $\mathbf{p}^{(t+1)}_{i}=\bfPhi_i\big(\bI_i^{(t)}\big)$, while all
other MUs keep their power allocations fixed. Although simple in
form, this algorithm requires additional overhead in order to
implement the sequential update. In addition, when the number of MUs
becomes large, such sequential algorithm typically converges slowly.
To overcome these drawbacks, we propose an alternative algorithm
that allows the MUs to update their power allocation {\it
simultaneously}.

{\bf Averaged Iterative-Water Filling Algorithm (A-IWF)}:\\
At each iteration $t$, the MUs do the following.

1) Calculate the WF power allocation
$\bfPhi_i\big(\bI^{(t)}_i\big),\ \forall~i\in\mathcal{N}$.

2) Adjust their power profiles simultaneously according to:{\small
\begin{align}
\mathbf{p}^{(t+1)}_{i}&=(1-\alpha^{(t)})\mathbf{p}^{(t)}_{i}+\alpha^{(t)}\bfPhi_i\big(\bI^{(t)}_i\big),
\ \forall \ i\in\cN
\end{align}}
where the sequence $\{\alpha^{(t)}\}_{t=1}^{\infty}$ satisfies
$\alpha^{(t)}\in(0,1)$ and :{\small
\begin{align}
\lim_{T\to\infty}\sum_{t=1}^{T}\alpha^{(t)}=\infty,
~\lim_{T\to\infty}\sum_{t=1}^{T}(\alpha^{(t)})^2<\infty.\label{eqAlphaProperty}
\end{align}}
The convergence property of the A-IWF algorithms is stated in the
following proposition, the proof of which can be found in Appendix
\ref{appProofAIWF}.

\newtheorem{P1}{Proposition}
\begin{P1}\label{propAIWF}
{\it Starting from any initial power
$\mathbf{p}^{(0)}\in\mathcal{P}$, the A-IWF algorithm converges to a
NE of game $G$.}
\end{P1}
We note that a similar algorithm has been recently proposed in
\cite{hong11a} for computing equilibria for a power control game in
the {\it interference channel}. However, the convergence of the
algorithm proposed in \cite{hong11a} requires some contraction
properties of the WF operator, which in turn requires that the
interference should be weak among the users. Unfortunately, such
stringent weak interference condition cannot be satisfied in our
current setting (see \cite{Mertikopoulos:2011:DPA:2151688.2151726}
for an argument). For this reason, completely different techniques
are used to prove the convergence of the A-IWF algorithm in the
present paper.

\vspace{-0.2cm}
\section{The Joint AP Selection and Power Control}\label{secSystemJoint}
In this section, we begin our discussion on the more general network
in the presence of multiple APs.

\vspace{-0.3cm}
\subsection{System Model}
Let us again index the MUs and the channels by the sets
$\mathcal{N}\triangleq\{1,2,\cdots,N\}$ and $\mathcal{K} \triangleq
\{1,2,\cdots,K\}$, respectively. Introduce the set $\mathcal{W}
\triangleq \{1,2,\cdots,W\}$ to index the APs. Again, we normalize
the total bandwidth to $1$. Each AP $w\in\mathcal{W}$ is assigned
with a subset of channels $\mathcal{K}_w\subseteq\mathcal{K}$. We
focus on the uplink scenario where each MU transmits to a single AP.
The
following are our main assumptions of the network.\\
{\bf A-1)} Each MU $i$ is able to associate to any AP.\\
{\bf A-2)} The APs operate on non-overlapping
portions of the available spectrum.\\
{\bf A-3)} Each AP is equipped with single-user receivers. 

Assumption A-1) is made merely for ease of presentation and
simplicity of notation. To account for the case that each MU has its
own set of candidate APs, one would simply need to introduce a
user-indexed subset of the APs as the feasible set of AP choices for
each MU.  Assumption A-2) is commonly used when considering AP
association problems in WLAN (e.g., see \cite{shakkottai07}), or the
vertical handoff problems in heterogenous networks (e.g., see
\cite{McNair04}), or the spectrum sharing problems in cognitive
network with multiple service providers (e.g., see
\cite{acharya09}). Assumption A-2) implies that
$\mathcal{K}_w\bigcap\mathcal{K}_q=\emptyset,~\forall~w\ne q,
k,q\in\mathcal{W}$.

We now list the additional notations needed for our analysis:
\begin{itemize}
\item $\{|h_{i,w}^k|^2\}_{k\in\mathcal{K}_w}$ and $\{n_w^k\}_{k\in\mathcal{K}_w}$ denote respectively, the power
gains from MU $i$ to AP $w$ on all channels and the thermal noise
powers on all channels for AP $w$.

\item $\mathbf{a} \in \mathcal{W}^{N}$ denotes
the {\it association profile} in the network, i.e. $\mathbf{a}[i]=w$
indicates that MU $i$ is associated to AP $w$.

\item $\mathcal{N}_w\triangleq \{i:\mathbf{a}[i]=w\}$ denotes the set of MUs that are associated
with AP $w$. By the restriction that each user can choose a single
AP, $\{\mathcal{N}_w\}_{w\in\mathcal{W}}$ is a partition of
$\mathcal{N}$.

\item ${p}_{i,w}^k$ represents the amount of power MU $i$ transmits on
channel $k$ when it is associated with AP $w$.

\item $\mathbf{p}_{i,w}\triangleq\big\{p_{i,w}^k\big\}_{k\in\mathcal{K}_w}$ \hspace{-0.2cm}denotes the
{power profile} of MU $i$ when $\ba[i]=w$;
$\mathbf{p}_{w}\triangleq\left\{\mathbf{p}_{i,w}\right\}_{i\in\mathcal{N}_w}$
denotes the power profiles of all MUs associated with AP $w$.

\item $\mathbf{p}_{-i,w}\triangleq\{\mathbf{p}_{j,w}\}_{j:j\ne i,
\mathbf{a}[j]=w}$ denotes the power profiles of all the MUs other
than $i$ that is associated with AP $w$.

\item $I_i^k\triangleq\sum_{j:\mathbf{a}[j]=w, j\ne
i}|h_{j,w}^k|^2p_{j,w}^k$ denotes the interference for MU $i$ on
channel $k$;
$\mathbf{I}_{i,w}\triangleq\left\{I_i^k\right\}_{k\in\mathcal{K}_w}$
is the set of all interferences on all channels of AP $w$.
\item {\small$\mathcal{F}_{i,w}\triangleq\Big\{\mathbf{p}_{i,w}:\hspace{-0.1cm}\sum_{k\in\mathcal{K}_w}p_{i,w}^k\le
P_i,~p_{i,w}^k\ge 0,~\forall~k\in\mathcal{K}_w\Big\}$} denotes MU
$i$'s feasible power allocation when it is associated with AP $w$.

\end{itemize}

When associated with AP $w$, MU $i$'s uplink transmission rate can
be expressed as\vspace{-0.15cm}{\small
\begin{align}
\hspace{-0.25cm}&R_i(\mathbf{p}_{i,w},\mathbf{p}_{-i,w};
w)=\frac{1}{K}\hspace{-0.15cm}\sum_{k\in\mathcal{K}_w}\hspace{-0.12cm}
\log\hspace{-0.08cm}\Bigg(\hspace{-0.05cm}1\hspace{-0.08cm}+
\hspace{-0.08cm}\frac{|h_{i,w}^k|^2p_{i,w}^k}{n_w^k+\sum_{j\ne i,\mathbf{a}[j]=w}|h_{j,w}^k|^2p_{j,w}^k}\Bigg)\label{eqTransmissionRate}\\
&\hspace{-0.25cm}=\frac{1}{K}\hspace{-0.12cm}\sum_{k\in\mathcal{K}_w}\hspace{-0.12cm}
\log\hspace{-0.08cm}\Bigg(\hspace{-0.05cm}1+\frac{|h_{i,w}^k|^2p_{i,w}^k}{n_w^k+I_{i}^k}\Bigg)\triangleq
R_i\big(\mathbf{p}_{i,w},\mathbf{I}_{i,w};w\big)\label{eqRAlternativeDefinition}.
\end{align}}
Fixing the association profile $\ba$ and all the other MUs' power
allocation $\bp_{-i,w}$, we use $\bfPhi_{i}\big(\bI_{i,w};w\big)$ to
denote the vector WF solution to MU $i$'s power optimization problem
$\max_{\bp_{i,w}\in\mathcal{F}_{i,w}}R_i(\mathbf{p}_{i,w},\mathbf{p}_{-i,w};w)$.



\vspace{-0.3cm}
\subsection{A Non-Cooperative Game Formulation}
We consider the non-cooperative game in which each MU selects an AP
and a power allocation over the channels at the selected AP. This
game can be formally described as follows: {\it i)} the MUs are the
players; {\it ii)} each MU's strategy space is $\chi_i\triangleq
\bigcup_{w\in\mathcal{W}}\left\{w, \mathcal{F}_{i,w}\right\}$; {\it
iii)} each MU's utility is $R_i(\mathbf{p}_{i,w},\mathbf{p}_{-i,w};
w)$. We denote this game as $
\mathcal{G}\triangleq\left\{\mathcal{N},\{{\chi}_i\}_{i\in\mathcal{N}},
\{R_i\}_{i\in\mathcal{N}}\right\}$. We emphasize that the feasible
strategy of a player contains a discrete variable and a continuous
vector, which makes the game $\mathcal{G}$ more complicated than
most of the games considered in the network resource allocation
literature.

The NE of this game is the tuple $\left\{\mathbf{a}^*[i],
\mathbf{p}^*_{i,{\mathbf{a}}^*[i]}\right\}_{i\in\mathcal{N}}$ that
satisfies the following set of equations:\vspace{-0.1cm}{\small
\begin{align}
\hspace{-0.1cm}\Big(\mathbf{a}^*[i],\mathbf{p}^{*}_{i,\mathbf{a}^*[i]}\Big)\in
\arg\max_{w\in\mathcal{W}}\max_{\bp_{i,w}\in\mathcal{F}_{i,w}}\hspace{-0.2cm}R_i(\mathbf{p}_{i,w},\mathbf{p}^{*}_{-i,w};
w),~\forall~i\label{eqNE}.
\end{align}}
We name the equilibrium profile $\mathbf{a}^*$  a {\it NE
association profile}, and
$\mathbf{p}^*({\mathbf{a}^*)}\triangleq\big\{\mathbf{p}^*_{i,{\mathbf{a}}^*[i]}\big\}_{i\in\mathcal{N}}$
a {\it NE power allocation profile}. To avoid duplicated
definitions, we name the tuple $\left(\mathbf{a}^*,
\mathbf{p}^*({\mathbf{a}^*)}\right)$ a {\it joint equilibrium
profile} (JEP) of the game $\mathcal{G}$ (instead of a NE). It is
clear from the above definitions that in a JEP, the system is stable
in the sense that no MU has the incentive to deviate from either its
AP association or its power allocation.

\vspace{-0.3cm} \subsection{Properties of the JEP}\label{subJEP}

In this section, we generalize the potential function for the power
allocation game ${G}$ to the game $\mathcal{G}$. We then prove that
the JEP always exists for the game $\mathcal{G}$.


We define the potential function for the game $\mathcal{G}$ as
follows.
\newtheorem{D1}{Definition}
\begin{D1}
{\it For a given tuple $(\ba, \bp)$, define the potential function
for the AP $w$ as :\vspace{-0.1cm}{\small
\begin{align}
P_w(\mathbf{p}_w;{\mathbf{a}})\triangleq\frac{1}{K}\sum_{k\in\mathcal{K}_w}\left(\log\Big(n_w^k+\sum_{i\in\mathcal{N}_w}|h_{i,w}^k|^2
p_{i,w}^k\Big)-\log n_w^k\right)\nonumber.\vspace{-0.3cm}
\end{align}}
Then the {potential function} for the game $\mathcal{G}$ is defined
as:
\begin{align}
P(\mathbf{p};\mathbf{a})\triangleq\sum_{w\in\mathcal{W}}P_w(\mathbf{p}_w;\mathbf{a}).
\end{align} }
\end{D1}
For a given association profile $\ba \in \mathcal{W}^{N}$, define
$\mathcal{F}_w(\mathbf{a})\triangleq
\prod_{i\in\mathcal{N}_w}\mathcal{F}_{i,w}$ as the joint feasible
set for the MUs in the set $\cN_w$. Let
$\mathcal{F}(\mathbf{a})\triangleq
\prod_{w\in\mathcal{W}}\mathcal{F}_{w}(\ba)$. Our next result
characterizes the system potential function. It is a straightforward
consequence of Theorem \ref{theoremPotential}, Corollary
\ref{corPotential}, and the observation that for a given association
profile $\ba$, the activities of two different MUs $i,j$ do not
affect each other as long as $\ba[i] \neq \ba[j]$.
\newtheorem{C2}{Corollary}
\begin{C1}\label{corollaryPotentialMaximization}
{\it For a given $\mathbf{a}$, a feasible
$\mathbf{p}^*\in\mathcal{F}(\mathbf{a})$ maximizes the potential
function $P(\mathbf{p};{\mathbf{a}})$ if and only if for all
$w\in\mathcal{W}$, $\bp^*_w$ maximizes the per-AP potential
function,
\begin{align}
\bp^*_w\in\max_{\bp_w\in\mathcal{F}_w(\ba)}P_w(\bp_w;\ba),\ \forall\
w\in\mathcal{W}.\label{eqSubProblemOptimization}
\end{align}
Moreover, $\bp^*_w$ is the NE for a single AP power allocation game
$G$, characterized by the potential function $P_w(\bp_w;\ba)$, and
with $\cN_w$ as the set of players.}
\end{C1}
We note that problem \eqref{eqSubProblemOptimization} is precisely
the single AP power allocation problem discussed in Section
\ref{secProblemFormulation}. Hence the power allocation $\bp^*_w$
can be computed distributedly by the set of MUs $\mathcal{N}_w$
using the A-IWF or S-IWF algorithm.

For a given $\ba \in \mathcal{W}^{N}$, let $\cE(\ba)$ and
$\bar{P}(\ba)$ denote the set of optimal solutions and the optimal
objective value of the problem
$\max_{\bp\in\mathcal{F}(\ba)}P(\bp;\ba)$, respectively. Similarly,
define $\cE_w(\ba)$ and $\bar{P}_w(\ba)$ as the set of optimal
solutions and the optimal objective value for problem
\eqref{eqSubProblemOptimization}. As discussed in Section
\ref{secProblemFormulation}, each subproblem in
\eqref{eqSubProblemOptimization} is a convex problem hence
$\bar{P}_w(\ba)$ is unique. It follows that for a given $\ba$,
$\bar{P}(\ba)$ takes a single value as well. Therefore,
$\bar{P}(\ba)$ can be viewed as a function of the association
profile $\ba$.

We emphasize that determining the existence of the JEP (which is a
{\it pure} NE) for the game $\mathcal{G}$ is by no means a trivial
proposition. Due to the mixed-integer structure of the game
$\mathcal{G}$, the standard results on the existence of the pure NE
of either continuous or discrete games cannot be applied.
Consequently, we have to explore the structure of the problem in
proving the existence of JEP for the game $\mathcal{G}$.

\newtheorem{T2}{Theorem}
\begin{T1}\label{theoremExistence}
The game $\mathcal{G}$ always admits a JEP in pure strategies. An
association profile
$\widetilde{\mathbf{a}}\in\arg\max_{\mathbf{a}}\bar{P}(\mathbf{a})$,
along with a power allocation profile $\btp=
\left\{\btp_{i,\widetilde{\mathbf{a}}[i]}\right\}_{i\in\mathcal{N}}\in\mathcal{E}(\widetilde{\mathbf{a}})$,
constitute a JEP of the game $\mathcal{G}$.
\end{T1}
\begin{proof}
We prove this theorem by contradiction. Suppose
$\widetilde{\mathbf{a}}$ maximizes the system potential, but
$\widetilde{\mathbf{a}}$ is not a NE association profile. Then there
must exist a MU $i$ who prefers to switch from
$\widetilde{\mathbf{a}}[i]=\widetilde{w}$ to a different AP
$\widehat{w}\ne \widetilde{w}$. Define a new profile
$\widehat{\mathbf{a}}=[\widehat{w}, \widetilde{\ba}_{-i}]$. Let
$\btp\in\mathcal{E}(\widetilde{\mathbf{a}})$ and
$\bhp\in\mathcal{E}(\widehat{\mathbf{a}})$. Suppose that all other
MUs do not change their actions, then the maximum rate that MU $i$
can obtain after switching to AP $\widehat{w}$ is given by {\small
\begin{align}
\widehat{R}_i\left(\bar{\mathbf{p}}_{i,\widehat{w}},\btp_{\widehat{w}};{\widehat{w}}\right)
&= \frac{1}{K}\sum_{k\in\mathcal{K}_{\widehat{w}}}
\log\left(\frac{n_{\widehat{w}}^k+\sum_{j:
\widetilde{\mathbf{a}}[j]=\widehat{w}}|h_{j,\widehat{w}}^k|^2
\tp^k_{j,\widehat{w}}+|h_{i,\widehat{w}}^k|^2
\bar{p}_{i,\widehat{w}}^k}{n_{\widehat{w}}^k+\sum_{j:
\widetilde{\mathbf{a}}[j]=\widehat{w}}|h_{j,\widehat{w}}^k|^2 \tp^k_{j,\widehat{w}}}\right)\nonumber\\
&=P_{\widehat{w}}(\bar{\mathbf{p}}_{i,\widehat{w}},\btp_{\widehat{w}};{\widehat{\mathbf{a}}})-
P_{\widehat{w}}(\btp_{\widehat{w}};{\widetilde{\mathbf{a}}})\label{eqEstimatedRate}
\end{align}}
where the vector $\bar{\mathbf{p}}_{i,\widehat{w}}$ is determined
by: $ \bar{\mathbf{p}}_{i,\widehat{w}}=
\arg\max_{\mathbf{p}_i\in\mathcal{F}_{i,\widehat{w}}}
\widehat{R}_i(\mathbf{p}_i,\btp_{\widehat{w}};{\widehat{w}}). $ The
rate
$\widehat{R}_i(\bar{\mathbf{p}}_{i,\widehat{w}},\btp_{\widehat{w}};{\widehat{w}})$
is MU $i$'s {\it estimate} of the maximum rate it can get if it were
to switch to AP $\widehat{w}$. Let
$R_i(\btp_{i,\widetilde{w}},\btp_{-i,\widetilde{w}};{\widetilde{w}})$
denote the {\it actual} transmission rate for MU $i$ in profile
$\widetilde{\mathbf{a}}$. Similarly as in \eqref{eqEstimatedRate},
it can be explicitly expressed as{\small
\begin{align}
&{R}_i(\btp_{i,\widetilde{w}},\btp_{-i,\widetilde{w}};{\widetilde{w}})=P_{\widetilde{w}}(\btp_{\widetilde{w}};{\widetilde{\mathbf{a}}})-
P_{\widetilde{w}}(\btp_{-i,\widetilde{w}};{\widetilde{\mathbf{a}}}).\label{eqEquilibriumRate}
\end{align}}
Because MU $i$ prefers $\widehat{w}$, from the definition of the JEP
\eqref{eqNE} it follows that its current rate must be strictly less
than its estimated maximum rate:{\small
\begin{align}
{R}_i(\btp_{i,\widetilde{w}},\btp_{-i,\widetilde{w}};{\widetilde{w}})<
\widehat{R}_i(\bar{\mathbf{p}}_{i,\widehat{w}},\btp_{\widehat{w}};{\widehat{w}}).
\label{eqRtLessThenRBarT+1}
\end{align}}
Combining \eqref{eqEstimatedRate},  \eqref{eqEquilibriumRate}
 and \eqref{eqRtLessThenRBarT+1} we conclude that:{\small
\begin{align}
P_{\widetilde{w}}(\btp_{\widetilde{w}};{\widetilde{\mathbf{a}}})-&
P_{\widetilde{w}}(\btp_{-i,\widetilde{w}};{\widetilde{\mathbf{a}}})
<P_{\widehat{w}}(\bar{\mathbf{p}}_{i,\widehat{w}},\btp_{\widehat{w}};{\widehat{\mathbf{a}}})-
P_{\widehat{w}}(\btp_{\widehat{w}};{\widetilde{\mathbf{a}}})\label{eqComparePotential1}.
\end{align}}
Note that the term
$P_{\widetilde{w}}(\btp_{-i,\widetilde{w}};{\widetilde{\mathbf{a}}})$
is equivalent to
$P_{\widetilde{w}}(\btp_{-i,\widetilde{w}};{\widehat{\mathbf{a}}})$
due to the equivalence of the following sets $ \{j:j\ne i,
\widetilde{\mathbf{a}}[j]=\widetilde{w}\}$ and $\{j:j\ne i,
\widehat{\mathbf{a}}[j]=\widetilde{w}\}$. Due to the assumption that
$\bhp\in\mathcal{E}(\bha)$, from Corollary
\ref{corollaryPotentialMaximization} we must have that
$\bhp_{\widetilde{w}}$ maximizes the potential function of AP
$\widetilde{w}$ for the given association profile $\bha$:
$\bhp_{\widetilde{w}}\in
\arg\max_{\mathbf{p}_{\widetilde{w}}\in\mathcal{F}_{\widetilde{w}}(\widehat{\mathbf{a}})}
P_{\widetilde{w}}(\mathbf{p}_{\widetilde{w}};{\widehat{\mathbf{a}}})$.
Using the fact  that the set of MUs associated with AP
$\widetilde{w}$ under profile $\widehat{\mathbf{a}}$ is the same as
the set of MUs associated with AP $\widetilde{w}$ under profile
$\widetilde{\mathbf{a}}$ excluding MU $i$, we must have
$\btp_{-i,\widetilde{w}}\in\mathcal{F}_{\widetilde{w}}({\widehat{\mathbf{a}}})$.
Consequently, the following is true:{\small
\begin{align}
 P_{\widetilde{w}}(\bhp_{\widetilde{w}};{\widehat{\mathbf{a}}})& \ge
 P_{\widetilde{w}}(\btp_{-i,\widetilde{w}};{\widehat{\mathbf{a}}})
 =P_{\widetilde{w}}(\btp_{-i,\widetilde{w}};\widetilde{\mathbf{a}})\label{eqPotentialOPT}
\end{align}}
Similarly, we have
 that:{\small
\begin{align}
P_{\widehat{w}}(\bhp_{\widehat{w}};{\widehat{\mathbf{a}}})\ge
P_{\widehat{w}}(\bar{\mathbf{p}}_{i,\widehat{w}},\btp_{\widehat{w}};{\widehat{\mathbf{a}}})
.\label{eqPotentialOPT+1}
\end{align}}
Combining \eqref{eqPotentialOPT}, \eqref{eqPotentialOPT+1} and
\eqref{eqComparePotential1}, we have that:{\small
\begin{align}
P_{\widetilde{w}}(\btp_{\widetilde{w}};{\widetilde{\mathbf{a}}})-
P_{\widetilde{w}}(\bhp_{\widetilde{w}};{\widehat{\mathbf{a}}})<
P_{\widehat{w}}(\bhp_{\widehat{w}};{\widehat{\mathbf{a}}})-
P_{\widehat{w}}(\btp_{\widehat{w}};{\widetilde{\mathbf{a}}})\nonumber
\end{align}}
Rearranging terms, we obtain: {\small
\begin{align}
P_{\widetilde{w}}(\btp_{\widetilde{w}};\widetilde{\mathbf{a}})+
P_{\widehat{w}}(\btp_{\widehat{w}}; \widetilde{\mathbf{a}})<
P_{\widehat{w}}(\bhp_{\widehat{w}};{\widehat{\mathbf{a}}})+
P_{\widetilde{w}}(\bhp_{\widetilde{w}};{\widehat{\mathbf{a}}}).
\label{eqTwoTermSumPotential}
\end{align}}
Finally, notice that $P_w(\bhp_w;\bha)=P_w(\btp_w;\bta)$ for all the
APs $w$ other than $\widetilde{w}$ and $\widehat{w}$. Then
\eqref{eqTwoTermSumPotential} is equivalent to: {\small\begin{align}
\sum_{w\in\mathcal{W}}P_{w}(\btp_{ {w}};{\widetilde{\mathbf{a}}})<
\sum_{w\in\mathcal{W}}P_w(\bhp_w;{\widehat{\mathbf{a}}})\label{eqSumPotentialInequality},
\end{align}}
\hspace{-0.2cm}which is further equivalent to: $
\bar{P}(\widetilde{\mathbf{a}})<\bar{P}(\widehat{\mathbf{a}})$. This
is a contradiction to the assumption that
$\bta\in\max_{\ba}\bar{P}(\ba)$.  We conclude that
$\widetilde{\mathbf{a}}$ must be a NE association profile. Clearly,
$\btp$ is a NE power allocation profile. Consequently,
$\left(\widetilde{\mathbf{a}}, \btp\right)$ is a JEP.
\end{proof}

\vspace{-0.3cm}
\section{The Proposed Algorithm}\label{secAlgorithmJEP}
In this section, we introduce our first algorithm, referred to as
the Joint Access Point Selection and Power Allocation (JASPA)
algorithm, that allows the MUs in the network to compute the JEP in
a distributed fashion. 

%
%

In the proposed algorithm, the AP selections and the power
allocation adjustments take place over different time scales. While
the APs are chosen in a relatively {\em slow} time scale, power
allocations are made in a {\em faster} time scale. The properly
chosen time scales enable the power control algorithm to reach an
equilibrium {before} the AP selections are updated. Once the
convergence of power allocation is achieved, the MUs attempt to find
different APs to further increase their rates. Let us assume that
each MU maintains a length $M$ memory, which operates in a first in
first out (FIFO) fashion. When MU $i$ decides that its next best AP
association should be $w_i^*$, it records the decision by a $W\times
1$ {\it best-reply vector}
$\mathbf{b}_i:\mathbf{b}_i=\mathbf{e}_{w^*_i}$, and pushes this
vector into its memory. In the next iteration, the actual AP
association decisions for the MUs are made 
based on the elements in their respective memories. Let
$(\bp^{(t)},\ba^{(t)})$ denote the power and association profile of
iteration $t$. The proposed algorithm is detailed as follows.

1) {\bf Initialization}: Let t=0, MUs randomly choose their APs.

2) {\bf Power Allocation}: Fix $\mathbf{a}^{(t)}$, in each cell, the
set of MUs $\mathcal{N}_w$ calculate the power allocations
$\bp^{(t+1)}_w\in\cE_w\big(\ba^{(t)}\big)$, either by A-IWF or S-IWF
algorithm. The process of reaching such intermediate equilibrium is
referred to as an ``inner loop".

3) {\bf Find the Best AP}: Each MU $i$ communicates with all the APs
in the network, obtains necessary information to find a set
$\mathcal{W}^{(t+1)}_i$ of APs such that all
$w\in\mathcal{W}^{(t+1)}_i$ satisfies:{
\begin{align}
\hspace{-0.5cm}\max_{\mathbf{p}_{i,w}\in\mathcal{F}_{i,w}}R_i\bigg(\mathbf{p}_{i,w},\mathbf{p}^{(t+1)}_{w};w\bigg)\ge
R_i\bigg(\mathbf{p}_{\mathbf{a}^{(t)}[i]}^{(t+1)};{\mathbf{a}^{(t)}[i]}\bigg).\label{eqBetterAP}
\vspace{-0.5cm}
\end{align}
} Randomly select the AP $w^*_i\in\mathcal{W}^{(t+1)}_i$. Set the
best-reply vector $\mathbf{b}^{(t+1)}_i=\mathbf{e}_{w^*_i}$.

4) {\bf Update the Probability Vector}: For each MU $i$, update a
$W\times 1$ probability vector $\bfbeta^{(t)}_i$ according
to:{\small
\begin{equation}
\label{eqUpdateBeta}
\begin{split}
\hspace{-0.2cm}\bfbeta^{(t+1)}_i=\left\{ \begin{array}{ll} \bfbeta
^{(t)}_i+\frac{1}{M}(\mathbf{b}^{(t+1)}_i-
\mathbf{b}^{(t-M)}_i)&~\textrm{if}~t\ge M\\
\bfbeta ^{(t)}_i+\frac{1}{M}(\mathbf{b}^{(t+1)}_i-
\mathbf{b}^{(1)}_i)&~\textrm{if}~M>t>0\\
\mathbf{b}^{(1)}_i &\textrm{if}~t=0. \\
\end{array} \right.
\end{split}
\end{equation}}
Remove $\mathbf{b}^{(t-M)}_i$ from the front of the memory if $t\ge
M$; push $\mathbf{b}_i^{(t+1)}$ into the end of the memory.

5) {\bf Find the Next AP}: Each MU $i$ samples the AP index
according to:  $\mathbf{a}^{(t+1)}[i]\sim {\rm
multi}(\bfbeta^{(t+1)}_i)$, where ${\rm multi(\cdot)}$ represents a
multinomial distribution.

6) {\bf Continue}: If $\mathbf{a}^{(t+1)}=\mathbf{a}^{(t+1-m)}$ for
all $m=1,\cdots, M$, stop. Otherwise let t=t+1, and go to Step 2).

%
 We are now ready to make several comments
regarding to the proposed JASPA algorithm.

\newtheorem{R1}{Remark}
\begin{R1}  It is crucial that each MU finally decides on choosing a
{\it single} AP. Failing to do so results in system instability, in
which the MUs switch AP association indefinitely. More precisely, it
is preferable that for all $i\in\mathcal{N}$, $
\lim_{t\to\infty}\bfbeta^{(t)}_i=\be_{w_i^*}$ for some  $w_i^*\in\mathcal{W}$. 
\end{R1}
\vspace{-0.1cm}

\newtheorem{R2}{Remark}
\begin{R1}
To perform step 3) of JASPA, MU $i$ needs to compute its WF solution
$\bfPhi_{i,w}(\bI^{(t+1)}_{i,w};w)$. For such purpose, only the
vector of total interference $\bI^{(t+1)}_{i,w}$ is needed from AP
$w$. This is precisely the necessary information needed for finding
the set $\mathcal{W}^{(t+1)}_i$ in step 3) of the algorithm.
\end{R1}

\begin{figure}[htb]
\vspace*{-.2cm} {\includegraphics[width=
0.8\linewidth]{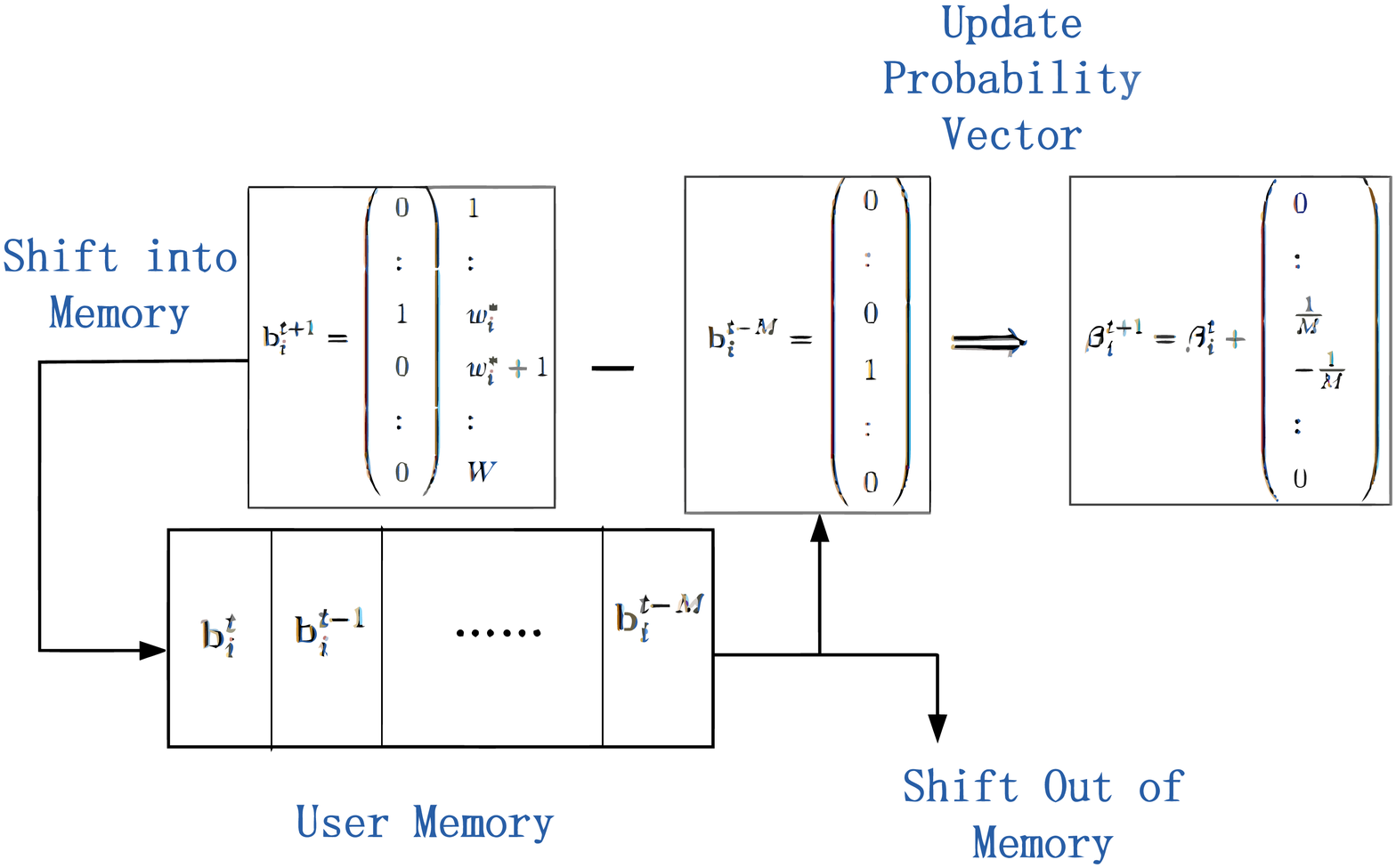}}
\vspace*{-.3cm}\caption{Graphical Illustration of Step 4 of the
JASPA algorithm. }\label{figMemory} \vspace*{-.5cm}
\end{figure}

\newtheorem{R5}{Remark}
\begin{R1}
In step 4) of JASPA, each MU's probability vector $\bfbeta_i$ is
updated. A graphical illustration of this step for user $i$ is
provided in Fig. \ref{figMemory}. By following the update rule in
\eqref{eqUpdateBeta}, the $w^*_i$-th entry of $\bfbeta_i$ will be at
least $\frac{1}{M}$ over the next $M$ iterations. This step combined
with the randomization procedure in step 5) ensure that all of MU
$i$'s best AP in the most recent $M$ iterations have the probability
of at least $\frac{1}{M}$ of being selected as $\ba^{(t+1)}[i]$.
This is in contrast to a naive greedy strategy that always selects
the best AP for the MUs in every iteration. In fact, such greedy
algorithm may diverge \footnote{Indeed, a simple example of the
divergence of the greedy algorithm is in a $2$ AP $2$ user network,
in which each AP has a single channel. Suppose the users' channels
are all identical, and both users associate with the same AP at the
beginning. Then each of them will always perceive the vacant AP as
its ``best AP". Each of them will then switch back and forth between
the APs and will never stabilize. }, while the proposed randomized
selection algorithm guarantees convergence (see Theorem
\ref{theoremConvergenceJASPA}.)
\end{R1}

\newtheorem{R3}{Remark}
\begin{R1}
To avoid unnecessary overhead related to ending and establishing the
connections, it is reasonable to assume that a selfish MU is willing
to leave its current AP only if the new AP can offer {\it
significant} improvement of the data rate. To model such behavior,
we introduce a {\it connection cost} $c_i\ge 0$ for each MU. Then
the set $\mathcal{W}_i^{(t+1)}$ should contain the APs that satisfy
the following inequality {\small
\begin{align}
\max_{\mathbf{p}_{i,w}\in\mathcal{F}_{i,w}}R_i\big(\mathbf{p}_{i,w},\mathbf{p}^{(t+1)}_{w};w\big)\ge
R_i(\mathbf{p}^{(t+1)};{\mathbf{a}^{(t)}[i]})+c_i.\nonumber
\end{align}}
From a system point of view, introducing these connection costs may
improve the convergence speed of the algorithm, as the MUs are less
willing to change associations. However it might also result in
reduced system throughput. 
\end{R1}

\newtheorem{R4}{Remark}
\begin{R1}\label{remarkTermination}
Suppose at time $T+1$, the algorithm terminates with profile
$\mathbf{a}^*$. Then $\mathbf{a}^*$ must be a NE association
profile. To see this, note that the algorithm stops when
$\mathbf{a}^*$ appears in $M$ {\it consecutive} iterations. From the
way that the {\it actual} AP association is generated, we see that
during iterations $(T-M+1,~T+1)$, each MU $i$ must prefer
$\mathbf{a}^*[i]$ at least once. More precisely, $\mathbf{a}^*$ must
be preferred by {all the MUs in the system}. It follows that
$\mathbf{a}^*$ is an equilibrium solution.
\end{R1}

The following theorem shows that the JASPA converges to a JEP {\it
globally} regardless of the starting points or the realizations of
the channel gains. See Appendix \ref{appProofJASPA} for proof.

\newtheorem{T6}{Theorem}
\begin{T1}\label{theoremConvergenceJASPA}
{\it When choosing $M\ge N$, the JASPA algorithm produces a sequence
$\left\{(\mathbf{a}^{(t)}, \mathbf{p}^{(t)})\right\}_{t=1}^{\infty}$
that converges to a JEP $(\mathbf{a}^*,\mathbf{p}^*(\mathbf{a}^*))$
with probability 1.}
\end{T1}
%

\vspace{-0.2cm}
\section{Extensions to the JASPA Algorithm}\label{secExtension}

The JASPA algorithm presented in the previous section is
``distributed" in the sense that the computation in each iteration
can be performed locally by the MUs. However, it requires the MUs to
jointly implement an {\it intermediate} power equilibrium
$\mathbf{p}^{(t+1)}$ between two AP selections $\ba^{(t)}$ and
$\ba^{(t+1)}$, which entails significant coordination among the MUs.
In this section, we propose two extensions of the JASPA algorithm
that do not require the MUs to reach any intermediate equilibria.

The first algorithm, named Se-JASPA, is a sequential version of the
JASPA. It is detailed in Table \ref{tableSeJASPA}.
\begin{table}[htb]
{\small
\begin{center}
\vspace{-0.1cm}
\begin{tabular}{|l|}
\hline 1) {\bf Initialization}: Each MU randomly chooses
$\mathbf{a}^{(0)}[i]$ and
$\mathbf{p}_{i,\mathbf{a}^{(0)}[i]}^{(0)}$\\
2) {\bf Determine the Next AP Association}:\\
\quad If it is MU $i$'s turn to act, (e.g., $\{(t+1) \textrm{mode} N
\}+1=i $):\\
\quad 2a) MU $i$ finds a set $ \mathcal{W}^{(t+1)}_i$ that satisfies\\
\quad\quad\quad $ \mathcal{W}^{(t+1)}_i= \{w^*:
\arg\max_{w\in\mathcal{W}}\max_{\mathbf{p}_{i,w}\in\mathcal{F}_{i,w}}
R(\mathbf{p}_{i,w},\mathbf{p}^{(t)}_{w};w)\}.
$\\

\quad 2b) MU $i$ selects an AP by randomly picking
$\mathbf{a}^{(t+1)}[i]\in\mathcal{W}_i^{(t+1)}$.\\
\quad For other MUs $j\ne i$, $\mathbf{a}^{(t+1)}[j]=\mathbf{a}^{(t)}[j]$.\\
3) {\bf Update the Power Allocation}: \\
\quad Denote $w_i^*=\mathbf{a}^{(t+1)}[i]$. MU $i$ updates by
$ \mathbf{p}^{(t+1)}_{i,w_i^*}=\bfPhi_i(\bI^{(t)}_{i,w_i^*};w_i^*)$. \\

\quad For other MUs $j\ne i$, $\mathbf{p}_{j,w}^{(t+1)}=\mathbf{p}_{j,w}^{(t)}$.\\

4) {\bf Continue}: Let t=t+1, and go to Step 2)\\

 \hline
\end{tabular}
\caption{The Se-JASPA Algorithm} \label{tableSeJASPA}
\end{center}
\vspace{-1cm} }
\end{table}

%

The Se-JASPA algorithm differs from the original JASPA algorithm in
several important ways. Firstly, each MU $i\in\cN$ does not need to
record the history of its best-reply vectors
$\{\mathbf{b}_i^{(t)}\}_{t}$. It decides on its AP association
greedily in step 2). Secondly, a MU $i$, after deciding a new AP
$\mathbf{a}^{(t+1)}[i]=w_i^*$, does not need to go through the
process of reaching an intermediate equilibrium. However, the MUs
still need to be coordinated for the exact sequence of their update,
because in each iteration only a single MU is allowed to act. As can
be inferred by the sequential nature of this algorithm, when the
number of MUs is large, the convergence becomes slow.

An alternative simultaneous version of the algorithm (named
Si-JASPA) is listed in Table \ref{tableSiJASPA}. Differently from
the Se-JASPA algorithm, it allows for all the users to update in
each iteration. We note that in the algorithm, the variable $T_i$
represents the duration that MU $i$ has stayed in the current AP;
the sequence of stepsizes $\{\alpha^{(t)}\}_{t=1}^{\infty}$ is
chosen according to \eqref{eqAlphaProperty}.

\begin{table}
{\small
\begin{center}
\vspace{-0.1cm}
\begin{tabular}{|l|}
\hline\\
1) {\bf Initialization (t=0)}:\\
\quad Each MU $i$ randomly chooses $\mathbf{a}^{(0)}[i]$ and
$\mathbf{p}_{i,\mathbf{a}^{(0)}[i]}^0$
\\
2) {\bf Selection of the Best Reply Association}:\\
\quad Each MU $i$ computes $\mathbf{b}^{(t+1)}_i$ following Step 3) of JASPA\\
3) {\bf Update Probability Vector}: \\
\quad Each MU $i$ updates $\bfbeta^{(t+1)}_i$ according
to \eqref{eqUpdateBeta} \\
\quad Push $\mathbf{b}_i^{(t+1)}$ into the memory; \\
\quad Remove $\mathbf{b}^{(t-M)}_i$ from the memory if $t\ge M$
\\
4) {\bf Determine the Next AP Association}: \\
\quad Each MU $i$ obtains $\mathbf{a}^{(t+1)}[i]$ following Step 5)
of JASPA
\\
5) {\bf Compute the Best Power Allocation}: \\
\quad Let $w^{*}_i=\mathbf{a}^{(t+1)}[i]$;\\
\quad Each MUs $i$ calculates
$\mathbf{p}^*_{i,w^*_i}$ by $\mathbf{p}^*_{i,w^*_i}=\bfPhi_i\big(\bI^{(t)}_{i,w_i^{*}};w_i^{*}\big)$\\
6) {\bf Update the Duration of Stay}: \\
\quad Each MU $i$ maintains and updates a variable $T_i$: \\
~~~~${T}_i=\left\{
\begin{array}{ll}
1&\textrm{if}~w^*_i\ne\mathbf{a}^{(t)}[i]\\
T_i+1&\textrm{if}~w^*_i=\mathbf{a}^{(t)}[i]\\
\end{array} \right.
$
\\
7) {\bf Update the Power Allocation}: \\
\quad Each MU $i$ calculates $\mathbf{p}^{(t+1)}_{i,w^*_i}$ as
follows:
\\
~~$\mathbf{p}^{(t+1)}_{i,w^{*}_i}=\left\{
\begin{array}{ll}
\mathbf{p}^*_{i,w^{*}_i}&~\textrm{if}~w^*_i\ne\mathbf{a}^{(t)}[i]\\
(1-\alpha^{(T_i)})\mathbf{p}^{(t)}_{i,w^{*}_i}+\alpha^{(T_i)}
\mathbf{p}^*_{i,w^{*}_i}&~~\textrm{if}~w^*_i=\mathbf{a}^{(t)}[i]\\
\end{array} \right.
$\\
8) {\bf Continue}: Let t=t+1, and go to Step 2)\\
 \hline
\end{tabular}
\caption{The Si-JASPA Algorithm} \label{tableSiJASPA}
\end{center}
\vspace{-1cm} }
\end{table}

The structure of the Si-JASPA is almost the same as the JASPA,
except that for each MU, after switching to a new AP, it does not
need to go through the process of joint computation of the
intermediate equilibrium. Instead, the MUs can make their AP
decisions in each iteration of the algorithm. The level of
coordination among the MUs required for this algorithm is minimum
among all the three algorithms introduced so far.

Although to this point there is no complete convergence results for
the Se/Si-JASPA algorithms, our simulations suggest that they indeed
converge. Moreover, the Si-JASPA usually converges faster than the
Se-JASPA.

\vspace{-0.3cm}
\section{JASPA Based on Network-Wide Joint-Strategy }
\label{secJJASPA}

In this section, an alternative algorithm with convergence guarantee
is proposed. This algorithm allows the MUs, as in the Se/Si-JASPA,
to jointly select their power profiles and AP associations without
the need to reach any intermediate equilibria. We will see later
that compared with the JASPA and its two variants introduced before,
the algorithm studied in this section requires considerably
different information/memory structure for the MUs and the APs.
Among others, it requires that the MUs maintain in their memory the
history of some network-wide {\it joint strategy} of all MUs. We
henceforth name this algorithm the Joint-strategy JASPA (J-JASPA).

\vspace{-0.3cm}
\subsection{The J-JASPA Algorithm}
The main idea of the J-JASPA algorithm is to let the MUs compute
their strategies according to some system state that is randomly
sampled from the history. This is different from the JASPA and
Si-JASPA algorithm, in which the MUs compute their best strategies
according to the {\it current} system state, but then perform their
actual association by random sampling the history of the best
strategies. To better introduce the proposed algorithm, we first
provide some definitions.
\begin{itemize}
\item Let $\mathcal{N}^{(t)}_w\triangleq\left\{i:\mathbf{a}^{(t)}[i]=w\right\}$
denotes the set of MUs that are associated with AP $w$ in iteration
$t$.

\item Let $\mathbf{I}(\mathcal{N}_w)\triangleq
\left\{\mathbf{I}_{i,w}\right\}_{i\in\mathcal{N}_w}$ be the joint
interference profile of the set of MUs $\mathcal{N}_w$.
\end{itemize}

We then elaborate on the required memory structure. Let each MU $i$
keep three different kinds of memories, each with length $M$ and
operates in a FIFO fashion.

\begin{enumerate}
\item The first memory, denoted as $\underline{\ba}_i$, records MU
$i$'s associated APs in the last $M$ iterations:
$\left\{\mathbf{a}^{(t)}[i]\right\}^{T}_{t=T-M+1}$, i.e.,
$\underline{\ba}_i[m]=\mathbf{a}^{(T-M+m)}[i]$, for $m=1,\cdots,M$.

\item The second memory, denoted as $\underline{\bI}_i$, records the MU
$i$'s interference levels in the last $M$ iterations,
$\left\{\mathbf{I}_i^{(t)}\right\}^{T}_{t=T-M+1}$, where
$\mathbf{I}^{(t)}_i\triangleq\{\mathbf{I}^{(t)}_{i,w}\}_{w\in\mathcal{W}}$.
That is, $\underline{\bI}_i[m]=\mathbf{I}_i^{(T-M+m)}$, for
$m=1,\cdots,M$.

\item The third memory, denoted as $\underline{\bR}_i$, records MU $i$'s rate in the last $M$ iterations,
$\left\{R_i\left(\mathbf{p}^{(t)}_{\mathbf{a}^{(t)}[i]};
\mathbf{a}^{(t)}[i]\right)\right\}^{T}_{t=T-M+1}$.
\end{enumerate}

Each AP $w$ is also required to keep track of some local quantities
\footnote{Here, ``local" means individual APs can gather these
pieces of information without the need to communicate with other
APs.} related to the history of the MUs' behaviors. Suppose a subset
of users $\mathcal{Q}\subseteq\mathcal{N}$ has been associated with
AP $w$ at least once during iterations $[0,~T]$. Let us define the
time index $t_w(\cQ)\triangleq\arg\max_{t}\{\cN^{(t)}_w=\cQ\}$ as
the most recent time index that $\cQ$ appears in AP $w$. The
following variables are required to be recorded by each AP
$w\in\mathcal{W}$:
\begin{itemize}
\item The local power  profile $\overline{\bp}_w(\mathcal{Q})=
\left\{\mathbf{p}^{{t_w}(\mathcal{Q})}_{i,w}\right\}_{i\in\mathcal{N}_w}$.
\item The local interference profile $\overline{\bI}_w(\mathcal{Q})=
\left\{\mathbf{I}^{{t_w}(\mathcal{Q})}_{i,w}\right\}_{i\in\mathcal{N}_w}$.
\item The total number of times that $\mathcal{Q}$ has appeared in AP $w$: $\overline{T}_w(\mathcal{Q})=\sum_{t\le
T}\mathbf{1}\{\mathcal{N}^{(t)}_w=\mathcal{Q}\}$, where
$\mathbf{1}\{\cdot\}$ is the indicator function.
\end{itemize}

The J-JASPA algorithm is stated as follows: \\
1) {\bf Initialization:} Let $t=0$, each MU $i$ randomly chooses
$\mathbf{a}^{(0)}[i]\in\mathcal{W}$ and
$\mathbf{p}^{(0)}_{i,\mathbf{a}^{(0)}[i]}\in\mathcal{F}_{i,\mathbf{a}^{(0)}[i]}$.\\
2) {\bf Update MU Memory:} For each $i\in\mathcal{N}$, obtain
$\mathbf{I}^{(t)}_i$ from the APs. If $t>M$, Remove
$\underline{\ba}_i[1]$, $\underline{\bI}_i[1]$ and
$\underline{\bR}_i[1]$. Push $\mathbf{a}^{(t)}[i]$,
$\mathbf{I}^{(t)}_i$, and
$R_i\left(\mathbf{p}^{(t)}_{\mathbf{a}^{(t)}[i]};
\mathbf{a}^{(t)}[i]\right)$ into the end of $\underline{\ba}_i$,
$\underline{\bI}_i$ and $\underline{\bR}_i$, respectively.
\\
3) {\bf Update AP Memory:} For each $w\in\mathcal{W}$, perform the
following updates:{\small
\begin{align}
&\overline{\bp}_w({\mathcal{N}^{(t)}_w})=\left\{\mathbf{p}^{(t)}_{i,w}\right\}_{i\in\mathcal{N}^{(t)}_w},\quad
\overline{\bI}_w({\mathcal{N}^{(t)}_w})=\left\{\mathbf{I}^{(t)}_{i,w}\right\}_{i\in\mathcal{N}^{(t)}_w},\nonumber\\
&\overline{T}_w({\mathcal{N}^{(t)}_w})=\overline{T}_w({\mathcal{N}^{(t)}_w})+1.\nonumber
\end{align}}\\
4) {\bf Sample Memory:} Let $\widehat{M}\triangleq\min\{M,t\}$, and
let $\be$ be the $\widehat{M}\times 1$ all $1$ vector.  Each MU $i$
samples its memories by:{\small
\begin{align}
&{\rm index}^{(t)}_i\sim {\rm
multi}\bigg(\frac{1}{\widehat{M}}\times \be\bigg);\nonumber
\end{align}}
Set $\widehat{{a}}^{(t)}_i=\underline{\ba}_i\big({\rm
index}^{(t)}_i\big)$,
$\widehat{\mathbf{I}}^{(t)}_i=\underline{\bI}_i\big({\rm
index}^{(t)}_i\big)$ and
$\widehat{R}^{(t)}_i=\underline{\bR}_i\big({\rm
index}^{(t)}_i\big)$.

5) {\bf The Best AP Association:} Each MU $i$ computes the set of
best APs $\mathcal{W}_i^*$ by the sampled variables
$\widehat{{a}}^{(t)}_i$, $\widehat{\mathbf{I}}^{(t)}_i$ and
$\widehat{R}^{(t)}_i$:{\small
\begin{align}
\mathcal{W}^*_i=
\left\{w:\max_{\mathbf{p}_{i,w}\in\mathcal{F}_{i,w}}
R_i\left(\mathbf{p}_{i,w},
\widehat{\mathbf{I}}^{(t)}_{i,w};w\right)>\widehat{R}^{(t)}_i\right\}\bigcup\widehat{a}^{(t)}_i.
\end{align}}
Randomly pick $w_i^*\in\mathcal{W}^*_i$, and set
$\mathbf{a}^{(t+1)}[i]=w_i^*$.\\
6) {\bf The Power Allocation:} Each MU $i$ switches to AP
$\mathbf{a}^{(t+1)}[i]=w^*_i$. MU $i$ obtains
$\overline{T}_{w^*_i}\big({\mathcal{N}^{(t+1)}_{w^*_i}}\big)$,
$\overline{\bI}_{w^*_i}\big({\mathcal{N}^{(t+1)}_{w^*_i}}\big)$, and
$\overline{\bp}_{w^*_i}\big({\mathcal{N}^{(t+1)}_{w^*_i}}\big)$ from
AP ${w^*_i}$.

 {\bf If}
$\overline{T}_{w^*_i}\big({\mathcal{N}^{(t+1)}_{w^*_i}}\big)\ge 1$,
then let
$\bar{t}_{w^*_i}=\overline{T}_{w^*_i}\big({\mathcal{N}^{(t+1)}_{w^*_i}}\big)$
and $\widehat{\alpha}=\alpha^{(\bar{t}_{w^*_i})}$. Choose power
according to{\small
\begin{align}
\mathbf{p}^{(t+1)}_{i,w^*_i}=(1-\widehat{\alpha})\overline{\bp}_{i,w^*_i}\big({\mathcal{N}^{(t+1)}_{w^*_i}}\big)+
\widehat{\alpha}\bfPhi_i\left(\bar{\mathbf{I}}_i\big({\mathcal{N}^{(t+1)}_{w^*_i}}\big);w^*_i\right).\label{eqJJASPAPowerUpdate}
\end{align}}
\quad {\bf Else} randomly pick
$\mathbf{p}^{(t+1)}_{i,w^*_i}\in\mathcal{F}_{i,{w^*_i}}$.\\
7) {\bf Continue:} Let $t=t+1$, go to step 2).


Comparing with the JASPA and Si-JASPA algorithms, the J-JASPA
algorithm possesses the following distinctive features.

\begin{enumerate}
\item In J-JASPA, each MU calculates its best AP association
according to a {\it sampled historical} network state, while in the
JASPA and Si-JASPA, it calculates this quantity according to the
{\it current} network state.

\item The J-JASPA algorithm requires the APs to have memory. Each AP needs
to record the local power allocation and interference profiles for
{\it all} the different sets of MUs that have been associated with
it in the previous iterations. No such requirement is imposed on all
the previously introduced algorithms.

\item  The J-JASPA requires larger memory for the MUs for constructing $\underline{\ba}_i$,
$\underline{\bI}_i$ and $\underline{\bR}_i$.

\item The J-JASPA requires extra communications between the MUs and
the APs. Such overhead mainly comes from  step 6), in which the MUs
retrieve information stored in the APs for their power updates.

\end{enumerate}


\vspace{-0.3cm}
\subsection{The convergence of the J-JASPA algorithm}
In this subsection, we show that the J-JASPA algorithm converges to
a JEP. The proofs of the claims made in this subsection can be found
in Appendix \ref{appProofJJASPASubsequenceConverge} to
\ref{appProofJJASPA}

Let a set $\mathcal{A}$ consist of association profiles that appear
infinitely often in the sequence $\{\ba^{(t)}\}_{t=1}^{\infty}$. We
first provide a result related to the power allocation along an
infinite subsequence in which $\bta\in\mathcal{A}$ appears.
\newtheorem{P7}{Proposition}
\begin{P1}\label{propIO2}
{\it Choose $\bta\in\mathcal{A}$. Let $\{t_n\}_{n=1}^{\infty}$ be a
subsequence such that $\left\{t_n:\ba^{(t_n)}=\bta\right\}$. Then
for all $w$, $\lim_{n\to\infty}\mathbf{p}_w^{(t_n)}=
\mathbf{p}_w^*$, which is an optimizer of the problem
$\max_{\bp_w\in\mathcal{F}_w(\bta)}P_w(\bp_w;\bta)$. Furthermore, we
have that $ \lim_{n\to\infty}P_w\left(\mathbf{p}_w^{(t_n)};
\mathbf{a}^{(t_n)}\right)=\bar{P}_w(\bta) $ and $
\lim_{n\to\infty}P\left(\mathbf{p}^{(t_n)},\ba^{(t_n)}\right)=\bar{P}(\bta).
$ }
\end{P1}

We need the following definitions to proceed. Let
$\widehat{\mathbf{a}}^{(t)}$ be the profile sampled from the memory
by the MUs in step 4) at time $t$:
$\widehat{\mathbf{a}}^{(t)}[i]\triangleq\widehat{a}^{(t)}_i,~\forall~i$.
For a specific $\bta$, let the subsequence
$\{\tilde{t}_n\}_{n=1}^{\infty}$ be the time instances that ${\bta}$
appears and is immediately sampled by all the MUs, i.e.,
$\left\{\tilde{t}_n: \ba^{(\tilde{t}_n)}=\bta ~\textrm{and}~
\widehat{\mathbf{a}}^{(\tilde{t}_n)}=\bta\right\}$. Note that if
$\mathbf{a}^{(t)}=\bta$, then according to step 3)-step 4),  with
non-zero probability, $\widehat{\mathbf{a}}^{(t)}=\bta$.  Thus, if
$\bta\in\mathcal{A}$, then $\{\tilde{t}_n\}$ is an infinite
sequence.

Define $R^*_i\left(
\widehat{\mathbf{I}}^{(t)}_{i,w};w\right)\triangleq\max_{\mathbf{p}_{i,w}\in\mathcal{F}_{i,w}}
R_i\left(\mathbf{p}_{i,w},
\widehat{\mathbf{I}}^{(t)}_{i,w};w\right)$ as the maximum rate MU
$i$ can achieve in AP $w$ based on the interference
$\widehat{\mathbf{I}}^{(t)}_{i,w}$. Define MU $i$'s best association
set as:{\small
\begin{align}
\cB_i\left(\widehat{\mathbf{I}}^{(t)}_i,\widehat{\mathbf{a}}^{(t)}[i]\right)\triangleq
\left\{w:R^*_i\left( \widehat{\mathbf{I}}^{(t)}_{i,w};w\right)>
\widehat{R}^{(t)}_i\right\}\bigcup\widehat{\mathbf{a}}^{(t)}[i]
\end{align}}
where $\widehat{R}^{(t)}_i$ is the sampled rate given in step 4).
From step 5) of the J-JASPA algorithm, all $w\in
\cB_i\left(\widehat{\mathbf{I}}^{(t)}_i,\widehat{\mathbf{a}}^{(t)}[i]\right)$
has non-zero probability to be the serving AP for MU $i$ in
iteration $t+1$. Let
${\mathbf{I}}^*_i(\bta)=\lim_{n\to\infty}\widehat{\mathbf{I}}^{(\tilde{t}_n)}_i$.
Due to Proposition \ref{propIO2}, and the fact that $\bta$ appears
infinitely often, such limit is well defined.
%
%
Next we provide an asymptotic characterization of the best
association set defined above.
\newtheorem{P8}{Proposition}
\begin{P1}\label{propInclusion}
{\it For a specific MU $i$ and a system association profile
$\bta\in\mathcal{A}$, suppose there exists a $w\ne\bta[i]$ such that
$w \in \cB_i\left({\mathbf{I}}^*_i(\bta),{\bta}[i]\right)$, i.e., MU
$i$ has the incentive to move to a different AP in the limit. Then
there exists a large enough constant $N_i^*(\bta)$ such that for all
$n> N_i^*(\bta)$, we have:{\small
\begin{align}
w\in
\cB_i\left(\widehat{\mathbf{I}}^{(\tilde{t}_n)}_i,\widehat{\mathbf{a}}^{(\tilde{t}_n)}[i]\right).\label{eqBestReplySetEqual}
\end{align}}}
\end{P1}
In words, when a profile $\bta$ appears infinitely often, and
suppose in the limit, when $\bta$ is sampled, a MU $i$ prefers
$w\ne\bta[i]$. Then it must prefer $w$ in every time instance
$\tilde{t}_n$ when $n$ is large enough. Now we are ready to provide
the main convergence result for the J-JASPA algorithm.

\newtheorem{T8}{Theorem}
\begin{T1}\label{theoremConvergenceJJASPA}
{\it When choosing $M\ge N$, the J-JASPA algorithm converges to a
JEP with probability 1.}
\end{T1}

\vspace{-0.3cm}
\section{Simulation Results}\label{secSimulation}

In this section, we demonstrate the performance of the JASPA
algorithm and its three variants discussed in this work. The
following simulation setting is considered. Multiple MUs and APs are
randomly placed in a $10 {\rm \ meter}$ by $10 {\rm \ meter}$ area.
We use $d_{i,w}$ to denote the distance between MU $i$ and AP $w$.
Unless otherwise noted, the channel gains between MU $i$ and AP $w$
are generated independently from an exponential distribution with
mean $\frac{1}{d^2_{i,w}}$ (i.e., $|h_{i,w}^k|$ is assumed to have
Rayleigh distribution). Pre-assign the available channels equally to
different APs. Throughout, a snapshot of the network refers to the
network with fixed (but randomly generated as above) AP, MU
locations and channel gains. The length of the individual memory is
set as $M=10$. For ease of comparison, when we use the JASPA
algorithm with connection cost, we set all the MUs' connection cost
$\{c_i\}_{i\in\mathcal{N}}$ to be identical.

We first show the results related to the convergence properties, and
then present the results related to the system throughput
performance. Due to the space limit, for each experiment we show the
results obtained by running either Si/Se-JASPA and J-JASPA, or those
obtained by the JASPA.

\subsubsection{Convergence} 
\begin{figure*}[ht] \vspace*{-.5cm}
    \begin{minipage}[t]{0.5\linewidth}
    \centering
    {\includegraphics[width=
1\linewidth]{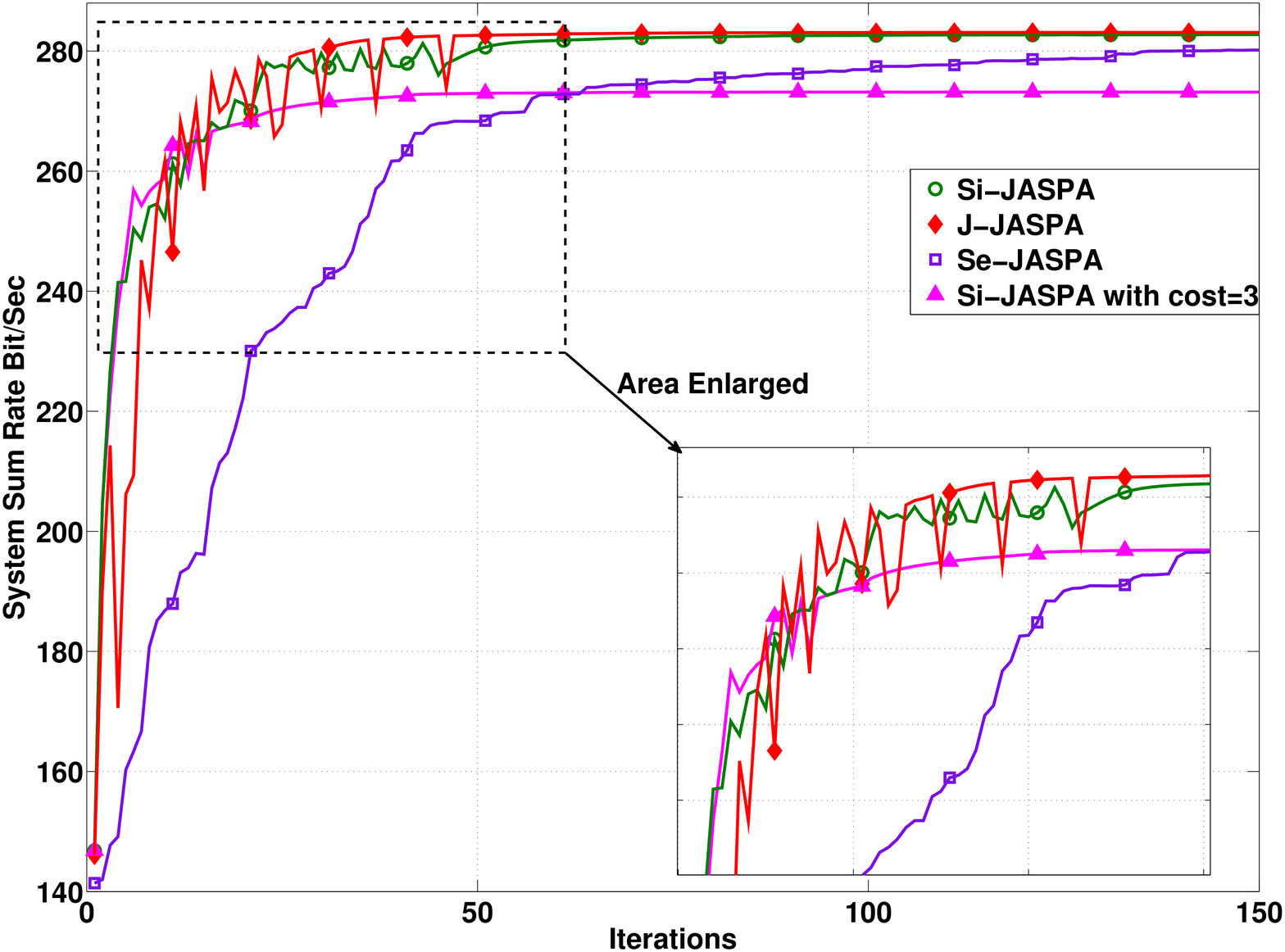} \vspace*{-0.5cm} \vspace*{-1cm}}
\end{minipage}
    \begin{minipage}[t]{0.5\linewidth}
    \centering
    {\includegraphics[width=
1\linewidth]{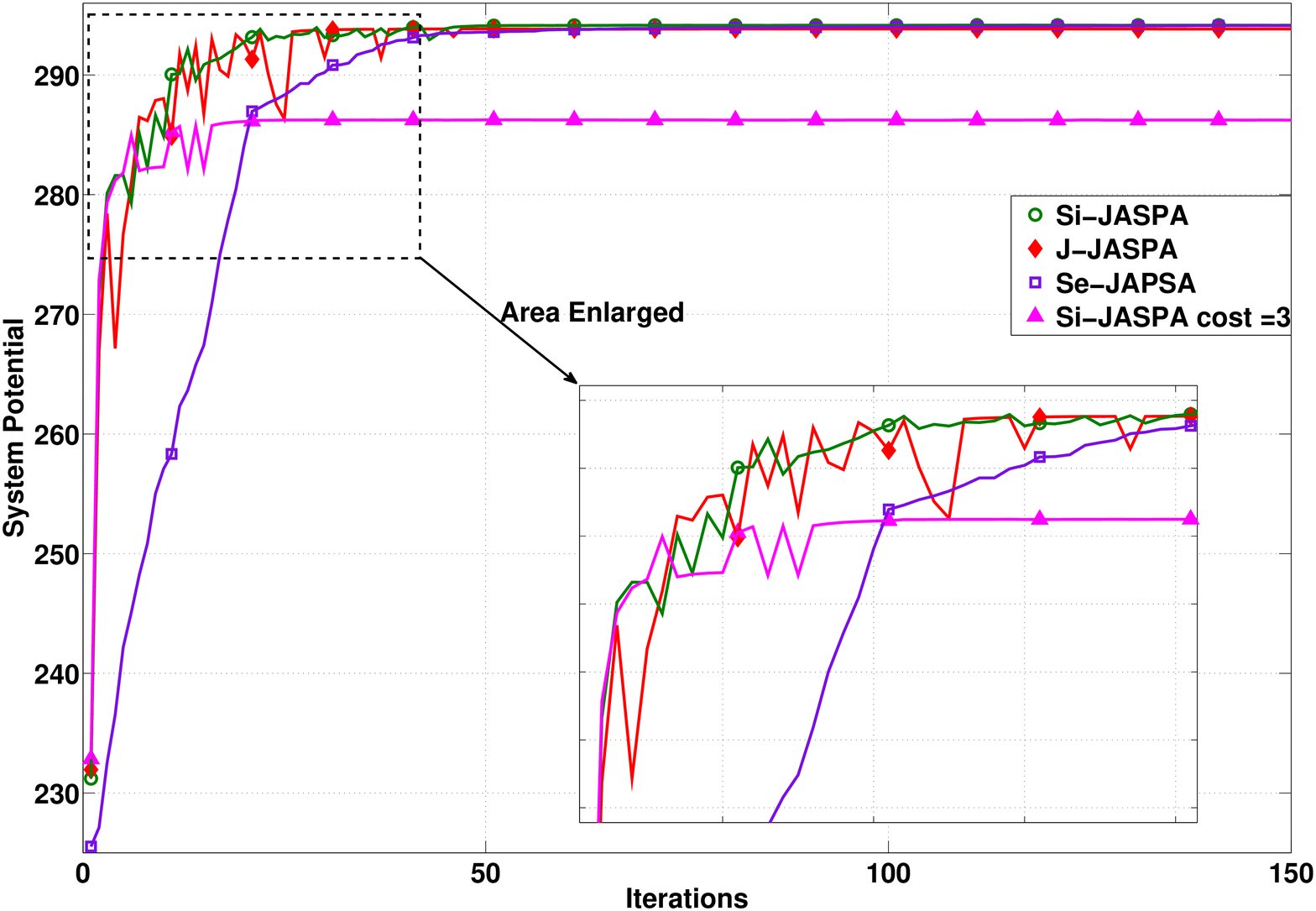} \vspace*{-0.5cm}
\vspace*{-1cm}}
\end{minipage}
\vspace{-0.3cm} \caption{Comparison of convergence speed by the
Si/Se-JASPA, J-JASPA and Si-JASPA with connection costs. $N=20$,
$K=64$, $Q=4$ and $c_i=3$ bit/s. Left: evolution of system sum rate.
Right: evolution of the system potential
function.}\label{figCompareSpeed}\vspace*{-0.2cm}
    \end{figure*}

Only the the results for Si/Se-JASPA
and J-JASPA are shown in this subsection. We first consider a
network with $20$ MUs, $64$ channels, and $4$ APs. Fig.
\ref{figCompareSpeed} shows the evolution of the system throughput
as well as the system potential function generated by a typical run
of the Se-JASPA, Si-JASPA, J-JASPA and Si-JASPA with connection cost
$c_i=3$ bit/sec $\forall~i\in\mathcal{N}$. We observe that the
Si-JASPA with connection cost converges faster than Si-JASPA and
Se-JASPA, while Se-JASPA converges very slowly. After convergence,
the system throughput achieved by Si-JASPA with connection cost is
smaller than those achieved by the other three algorithms.

   \begin{figure*}[htb] \vspace*{-.2cm}
        \begin{minipage}[t]{0.33\linewidth}
    \centering
    {\includegraphics[width=
1\linewidth]{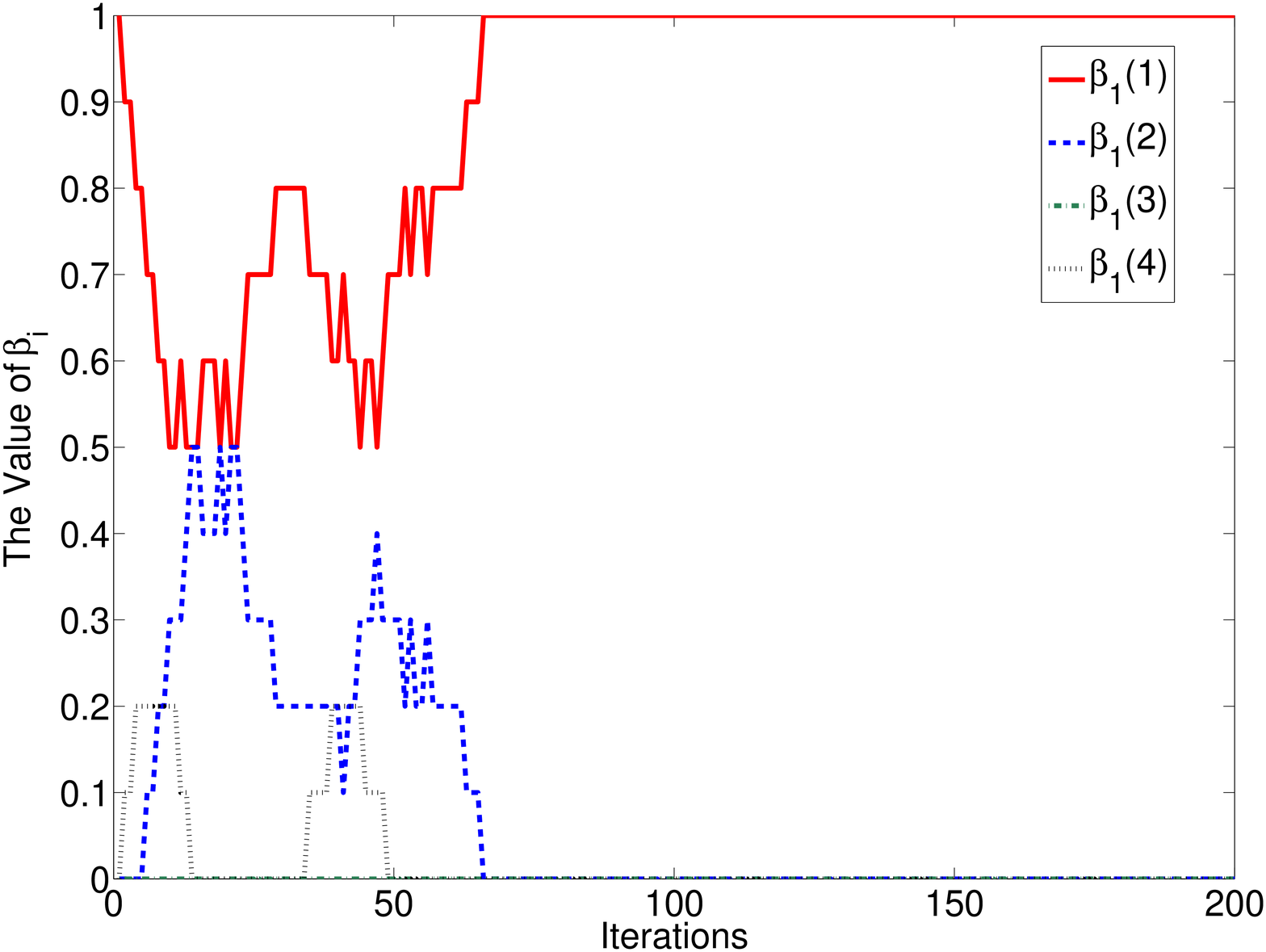}
\vspace*{-.4cm}}
\end{minipage}
\begin{minipage}[t]{0.33\linewidth}
    \centering
    {\includegraphics[width=
1\linewidth]{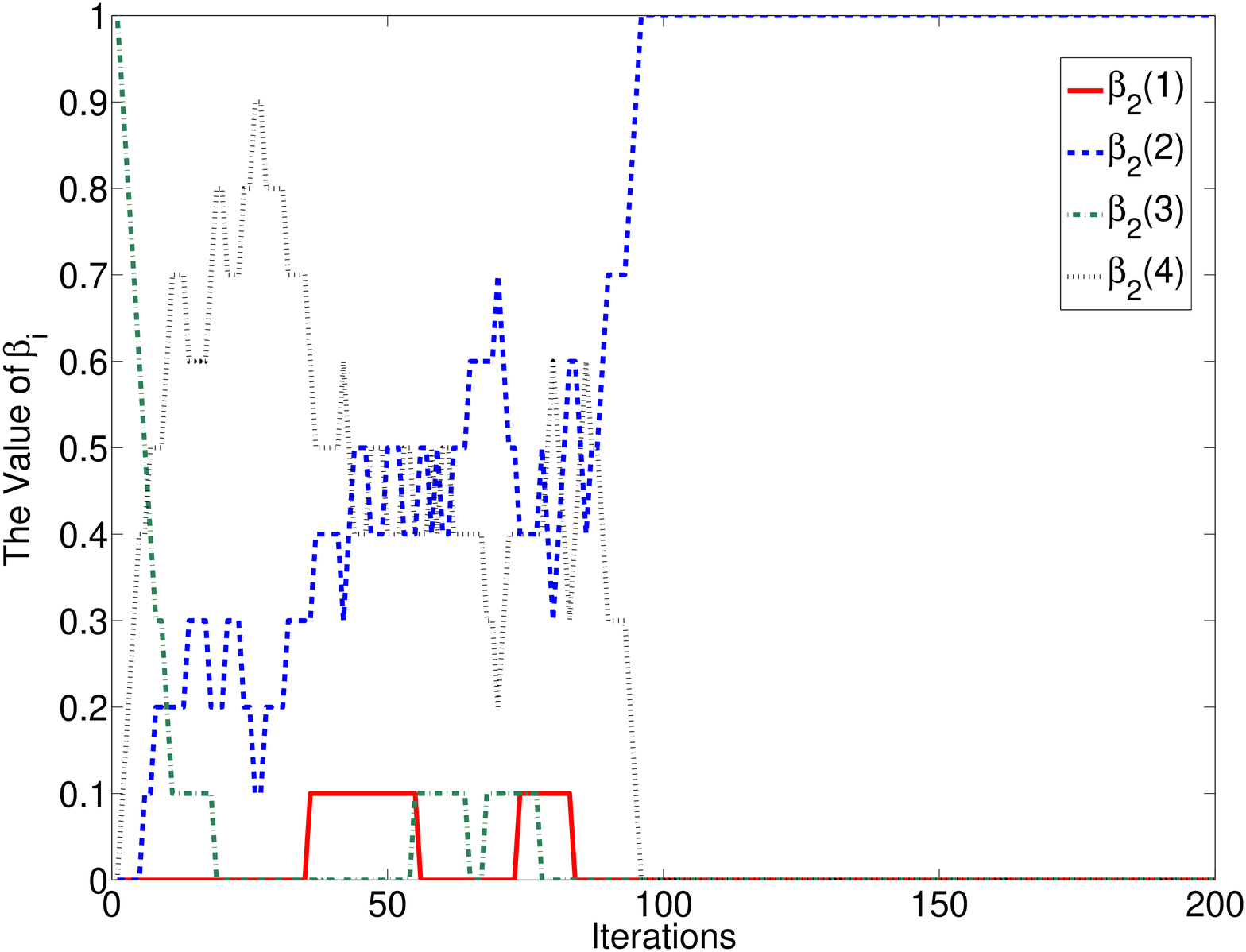}
\vspace*{-.4cm}}
\end{minipage}
    \begin{minipage}[t]{0.33\linewidth}
    \centering
    {\includegraphics[width=
1\linewidth]{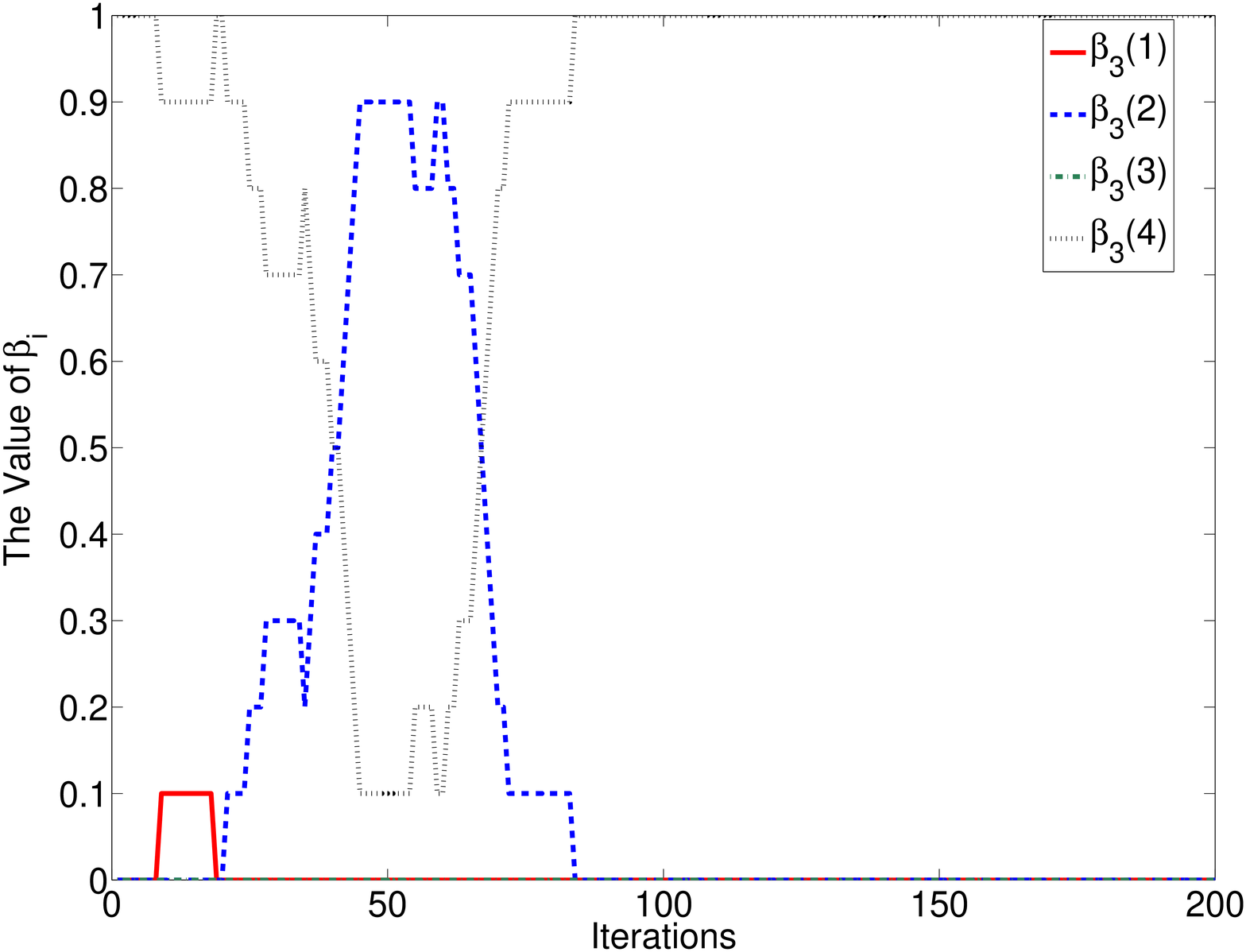}
\vspace*{-0.4cm}}
\end{minipage}
\caption{Convergence of $\beta^{(t)}_1$, $\beta^{(t)}_2$ and
$\beta^{(t)}_3$ generated by a typical run of Si-JASPA. $N=20$,
$K=64$ and $Q=4$. }\label{figConvergenceBeta}\vspace*{-0.4cm}
    \end{figure*}

\begin{figure*}[htb] \vspace*{-.2cm}

    \begin{minipage}[htb]{0.48\linewidth}
    \centering
    {\includegraphics[width=
1\linewidth]{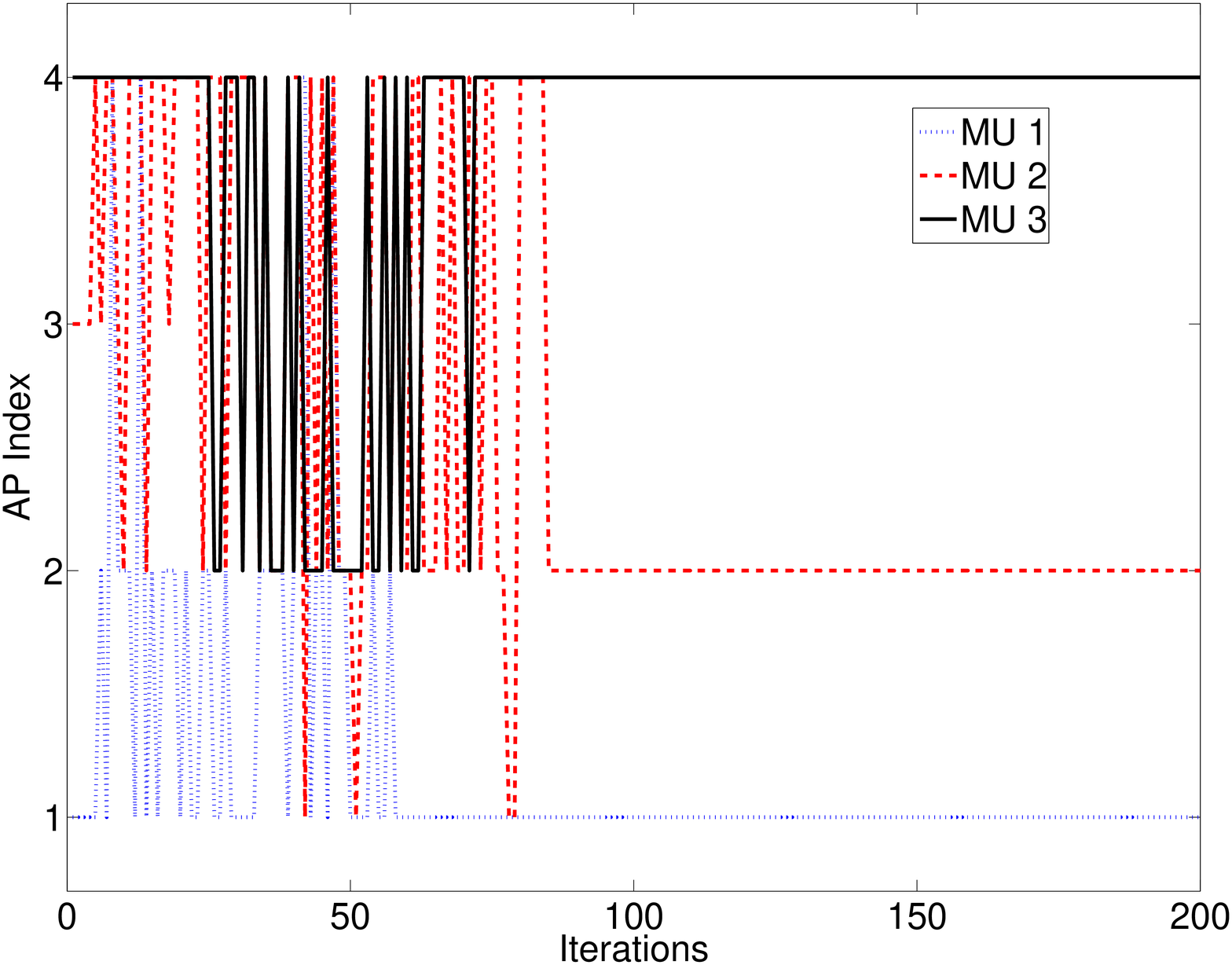}
\vspace*{-0.7cm}\caption{Convergence of selected MUs' AP selection
generated by a typical run of Si-JASPA. $N=20$, $K=64$ and $Q=4$.
}\label{figConvergenceSelection} \vspace*{-1cm}}
\end{minipage}\hfill
    \begin{minipage}[htb]{0.48\linewidth}
    \centering
    {\includegraphics[width=
1\linewidth]{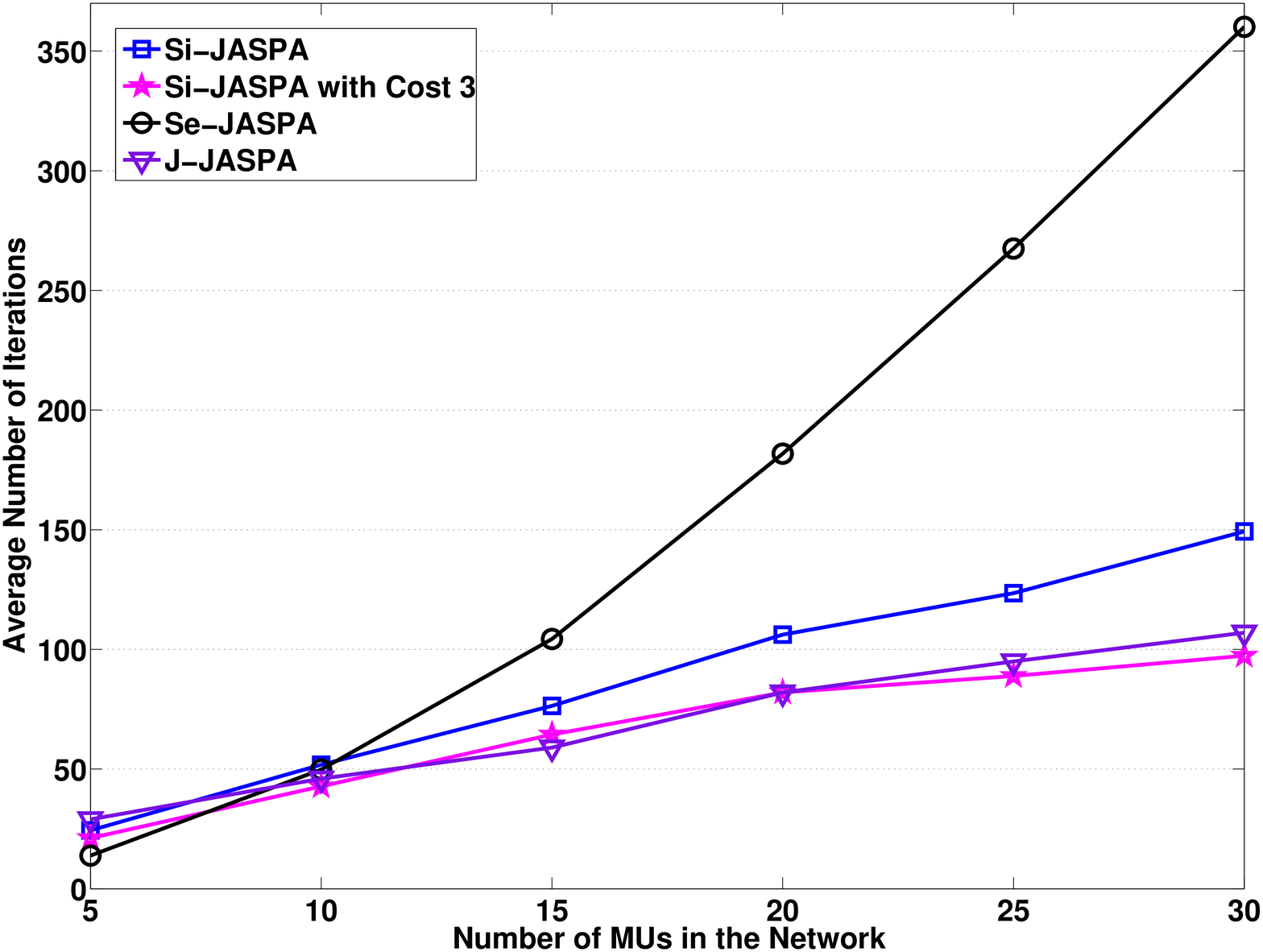}
\vspace*{-0.7cm}\caption{The averaged number of iterations required
for the convergence for different algorithms. $K=64$, $Q=4$, and
$c_i=3$ bit/s. Each point in figure is an average of $100$
independent runs of the algorithms. }\label{figConvergenceSpeed}
\vspace*{-0.4cm}}
\end{minipage}
\end{figure*}

Fig. \ref{figConvergenceSelection} shows the evolution of the AP
selections made by the MUs in the network during a typical run of
the Si-JASPA algorithm. We only show 3 out of 20 MUs (the selected
MUs are labeled as MU 1, 2, 3 for easy reference) in order not to
make the figure overly crowded. Fig. \ref{figConvergenceBeta} shows
the corresponding evolution of the probability vectors
$\{\bfbeta^{(t)}_i\}_{t=1}^{200}$ for the three of the MUs selected
in Fig. \ref{figConvergenceSelection}. It is clear that upon
convergence, all the probability vectors converge  to unit vectors.
Fig. \ref{figConvergenceSpeed} evaluates the impact of the number of
MUs $N$ on the speed of convergence of different algorithms. When
$N$ becomes large, the sequential version of the JASPA takes
significantly longer time to converge than the other three
simultaneous versions of the algorithm. Moreover, the J-JASPA
exhibits faster convergence than the Si/Se-JASPA. It is also
noteworthy to mention that including the connection costs indeed
helps to accelerate the convergence for the algorithm.

\vspace{-0.2cm}
\subsubsection{System Throughput Performance}\label{subThroughput}

We evaluate the throughput achievable by the JEP computed by the
JASPA. Such throughput performance is compared against a simple
baseline algorithm that assigns the users to their closest AP in
terms of actual distance. After assignment, the power allocation is
computed using the A-IWF algorithm discussed in Section
\ref{secProblemFormulation}.

We first consider a small network with $8$ MUs, $64$ channels and
$W\in\{1,~2,~3,~4\}$ APs. We compare the performance of the JASPA
algorithms to the maximum throughput that can be achieved. The
maximum throughput for a snapshot of the network is calculated by an
exhaustive search procedure: 1) for a given association profile, say
$\mathbf{a}$, calculate the maximum throughput (denoted by
$T(\mathbf{a})$) by summing up the maximum achievable rate of all
APs in the network; 2) enumerate {\it all possible} association
profiles, and find $T^*=\max_{\mathbf{a}}T(\mathbf{a})$. 

\begin{figure*}[htb] \vspace*{-.6cm}
    \begin{minipage}[htb]{0.48\linewidth}
    \centering
    {\includegraphics[width=
1\linewidth]{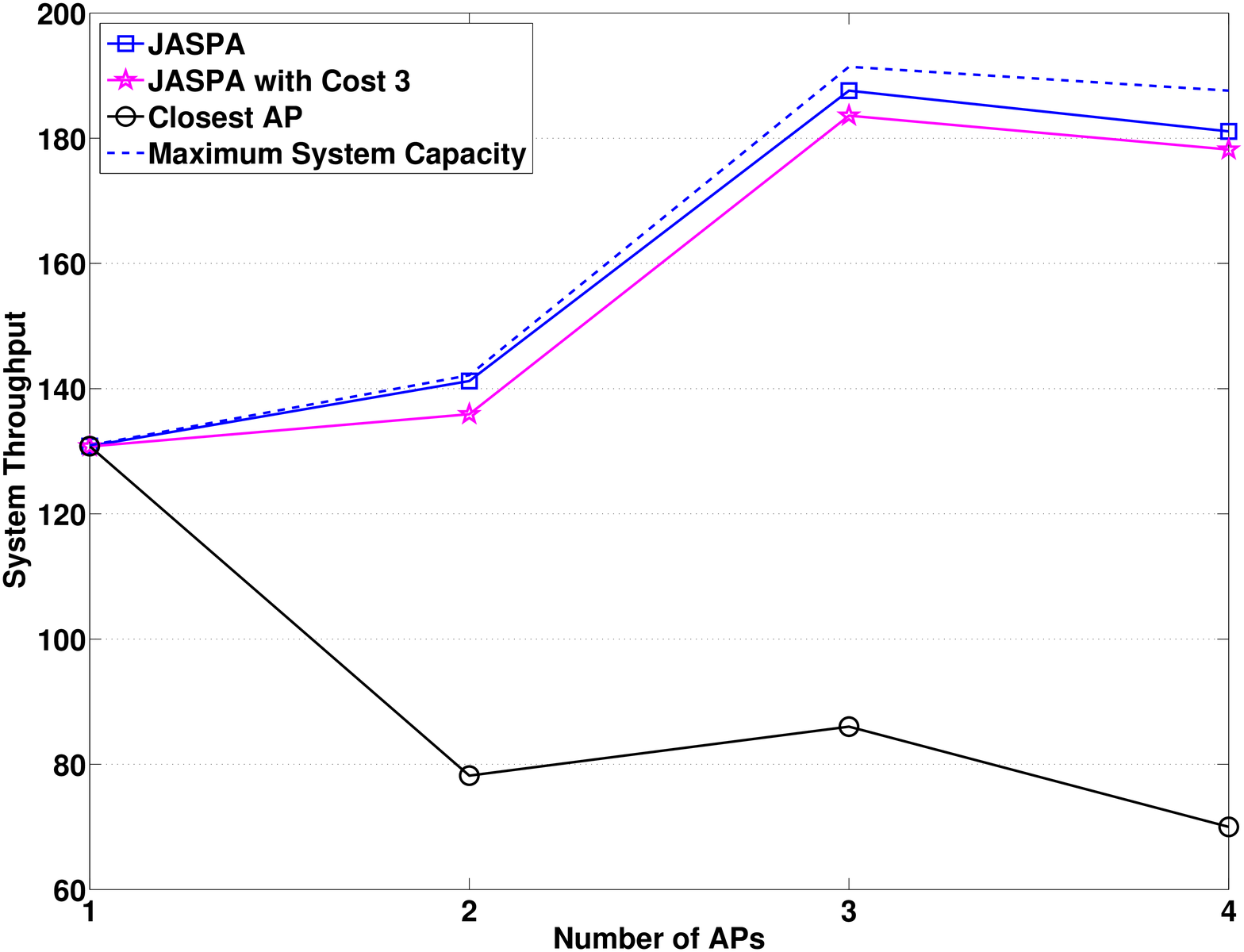}
\vspace*{-0.5cm}\caption{Comparison of the system throughput by
different algorithms. $ N= 8$, $K=64$ and $Q\in\{1,2,3,4\}$. Each
point in figure is obtained by running the algorithm on $100$
snapshots of the network.}\label{figCompareThrouputSmallScale}
\vspace*{-0.6cm}}
\end{minipage}\hfill
    \begin{minipage}[htb]{0.48\linewidth}
    \centering
    {\includegraphics[width=
1\linewidth]{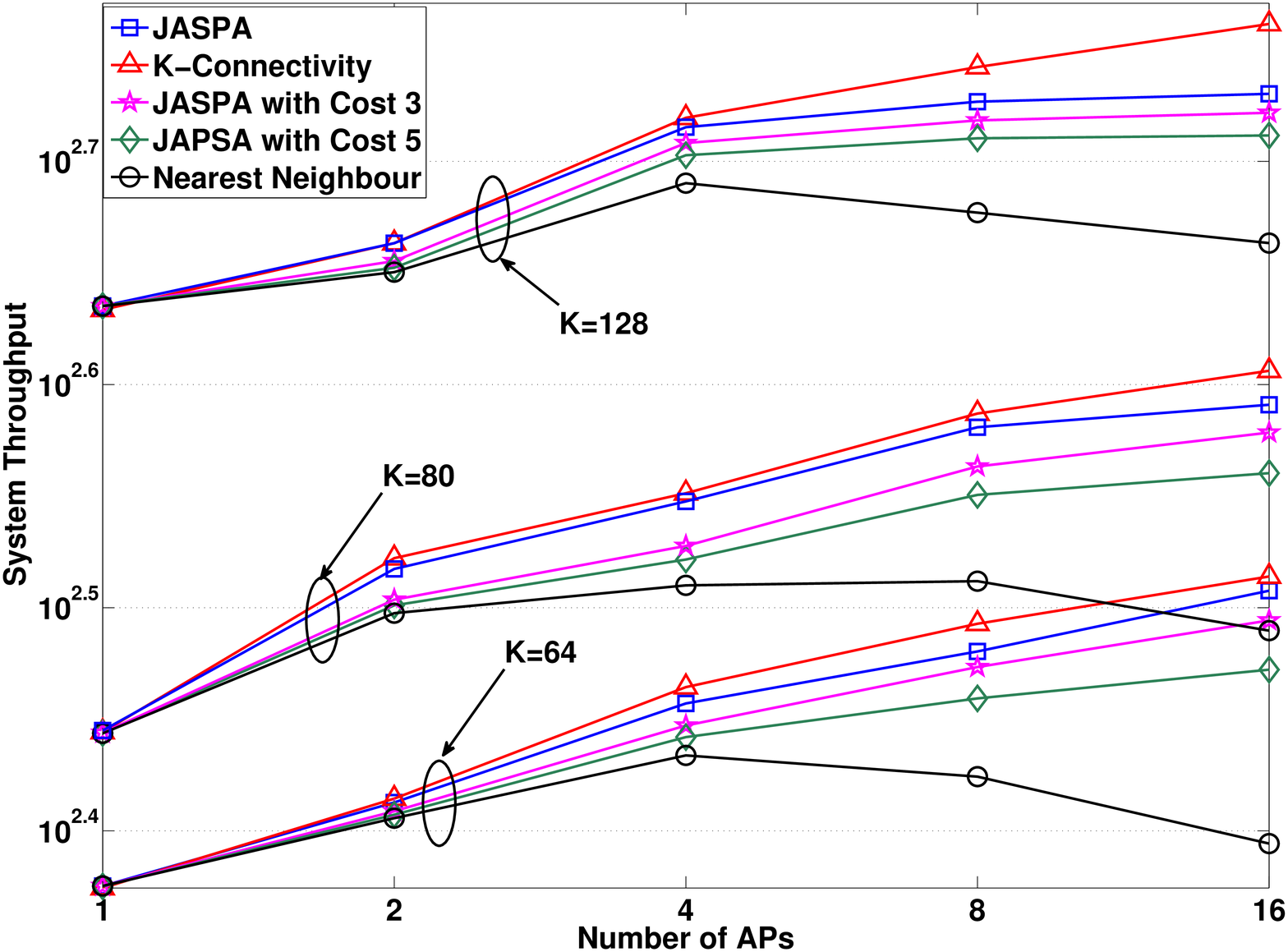}
\vspace*{-0.5cm}\caption{Comparison of the system throughput. $ N=
30$, $K\in\{64, 80, 128\}$ and $Q\in\{1, 2, 4, 8, 16\}$. Each point
in figure is obtained by running the algorithm on $100$ snapshots of
the network.}\label{figCompareThrouput} \vspace*{-0.6cm}}
\end{minipage}
    \end{figure*}

In Fig. \ref{figCompareThrouputSmallScale}, we see that the JASPA
algorithm performs very well with little throughput loss, while the
closest AP algorithm performs poorly.  We then investigate the
performance of larger networks with 30 MUs, up to 16 APs and up to
128 channels. Fig. \ref{figCompareThrouput} shows the comparison of
the performance of JASPA, JASPA with individual cost $c_i=3$ bit/sec
and $c_i=5$ bit/sec.

Due to the prohibitive computation time required, we are unable to
obtain the maximum system throughput for these relatively large
networks. For comparison purpose, we instead compute the equilibrium
system throughput that can be achieved in a game {\it if all the MUs
are able to connect to all the APs at the same time}. In such ideal
network, all the APs are pooled together as a single ``virtual" AP,
which along with the users form a ``virtual" MAC channel. The users
can allocate their transmit power by computing the NE for the
corresponding single AP game, as discussed in Section
\ref{secProblemFormulation}.  As suggested in Section
\ref{subSingleAPGame}, when the number of channel becomes large, the
throughput achieved by the NE of this power allocation game achieves
the capacity of the ``virtual" MAC. However, we observe that the
performance of JASPA is close to that of such ideal
``multiple-connectivity" network.


\vspace{-0.3cm}
\section{Conclusion}\label{secConclusion}
In this paper, we addressed the joint AP selection and power
allocation problem in a multichannel wireless network. The problem
was formulated as a non-cooperative game with mixed-integer strategy
space. We characterized the NEs of this game, and provided
distributed algorithms to reach the NEs. Empirical evidence gathered
from simulation suggests that the quality of the equilibrium
solutions is reasonably high.

There can be many future extensions to this work. First of all, the
non-cooperative game with mixed-integer strategy space analyzed in
this paper can be applied to many other problems as well, for
example, the {\it downlink} counterpart of the current problem.
Secondly, for the problem considered in this work, it is beneficial
to characterize quantitatively the efficiency of the JEP, and to
provide solutions for efficiency improvement. Thirdly, it would be
interesting to investigate the effect of time-varying channels and
the arrival and departure of the MUs on the performance of the
algorithm.

\vspace{-0.3cm}
\section{Appendix}
\vspace{-0.2cm}
\subsection{Proof of Proposition \ref{propAIWF}}\label{appProofAIWF}
Note that the WF operator $\bfPhi_i(\bI_i)$ is also a function of
$\bp_{-i}$, hence we can rewrite it as $\bfPhi_i(\bp_{-i})$. Define
$\bfPhi(\mathbf{p})\triangleq[\bfPhi_1(\mathbf{p}_{-1}),\cdots,\bfPhi_N(\mathbf{p}_{-N})]^{\intercal}$.
Define
$\mathbf{s}(\mathbf{p})\triangleq\bfPhi(\mathbf{p})-\mathbf{p}$.
Then the A-IWF algorithm can be expressed as:
\begin{align}
\mathbf{p}^{(t+1)}=(1-\alpha^{(t)})\mathbf{p}^{(t)}+\alpha^{(t)}
\bfPhi(\mathbf{p}^{(t)})=\mathbf{p}^{(t)}+\alpha^{(t)}\mathbf{s}(\mathbf{p}^{(t)}).\nonumber
\end{align}

We first state and prove two lemmas.

\newtheorem{L1}{Lemma}
\begin{L1}\label{lemmaAscend}
{\it We have $
\mathbf{s}(\mathbf{p})^{\intercal}\triangledown_{\mathbf{p}}P(\mathbf{p})\ge
M \|\mathbf{s}(\mathbf{p})\|^2, \|\bs(\bp)\|\le
Q\|\triangledown_{\mathbf{p}}P(\mathbf{p})\| $, where $M>0$ and
$Q>0$ are two constants.}
\end{L1}

\begin{proof}
To prove the first inequality, we need to show that:{\small
\begin{align}
\sum_{i=1}^{N}\sum_{k=1}^{K}\triangledown_{p_i^k}P(\mathbf{p})s^k_i(\mathbf{p})\ge
\sum_{i=1}^{N}\sum_{k=1}^{K}M\left(s^k_i(\mathbf{p})\right)^2
\label{eqM}
\end{align}}
where $s^k_i(\mathbf{p})\triangleq \Phi^k_i(\mathbf{p}_{-i})-p^k_i$.
It suffices to show that for each $i\in\mathcal{N}$, there exists
$M_i>0$ such that: 
{\small
\begin{align}
\sum_{k=1}^{K}\left(\triangledown_{p^k_i}P(\mathbf{p})-M_i
s^k_i(\mathbf{p})\right)s^k_i(\mathbf{p})\ge 0,\
\forall~i\in\cN.\label{eqPMinorS}
\end{align}}
Expressing $\triangledown_{p^k_i}P(\mathbf{p})$ explicitly, we
have{\small
\begin{align}
\triangledown_{p^k_i}P(\mathbf{p})&=\frac{|h^k_i|^2}{n^k+\sum_{j=1}^{N}|h^k_j|^2p^k_j}>
0.\label{eqPotentialGradient}
\end{align}}
Note that the Lagrangian multiplier $\sigma_i$ ensures the tightness
of MU $i$'s power budget constraint. Therefore we must have
$\sum_{k=1}^{K}p^k_{i}=P_i$, which in turn implies:{\small
\begin{align}
\sum_{k=1}^{K}s^k_i(\mathbf{p})=\sum_{k=1}^{K}\Phi^k_i(\mathbf{p}_{-i})-\sum_{k=1}^{K}p^k_i
=P_i-P_i=0. \label{eqSumSZero}
\end{align}}
Define two index sets $\cK_1\triangleq\{k: s^k_i(\bp)\ge0\}$,
$\cK_2\triangleq\{k: s^k_i(\bp)<0\}$. Then each inequality in
\eqref{eqPMinorS} is equivalent to {\small
\begin{align}
&\sum_{k\in \cK_1}\left(\triangledown_{p^k_i}P(\mathbf{p})-M_i
s^k_i(\mathbf{p})\right)s^k_i(\mathbf{p})\nonumber\\
&\ge \sum_{k\in \cK_2}\left(\triangledown_{p^k_i}P(\mathbf{p})-M_i
s^k_i(\mathbf{p})\right)(-s^k_i(\mathbf{p}))
\label{eqPMinorSEquivalent}.
\end{align}}
From \eqref{eqSumSZero} we have
$\sum_{k\in\cK_1}s^k_i(\bp)=\sum_{k\in\cK_2}(-s^k_i(\bp))$. Using
this result, we see that to show \eqref{eqPMinorSEquivalent}, it
suffices to show that there exists $M_i>0$ such that:{\small
\begin{align}
\hspace{-0.2cm}\min_{k\in\cK_1}\{\triangledown_{p^k_i}P(\mathbf{p})-M_i
s^k_i(\mathbf{p})\}\ge\max_{k\in\cK_2}\{\triangledown_{p^k_i}P(\mathbf{p})-M_i
s^k_i(\mathbf{p})\}. \label{eqPMinorSEquivalentMaxMin}
\end{align}}
Pick any $k_1\in\cK_1$, $k_2\in\cK_2$. Below we will show that for
the pair $(k_1,k_2)$ there exists $M^{(k_1,k_2)}_i >0$ such
that:
{\small
\begin{align}
\hspace{-0.2cm}\triangledown_{p_i^{k_1}}P(\mathbf{p})-\triangledown_{p_i^{k_2}}P(\mathbf{p})\ge
M^{(k1,k2)}_i \left(s^{k_1}_i(\mathbf{p})-
s^{k_2}_i(\mathbf{p})\right)\label{eqK1K2Inequality}.
\end{align}}
Note that if \eqref{eqK1K2Inequality} is true, we can take
$M_i=\min_{k_1,k_2}M^{(k_1,k_2)}_i$, then
\eqref{eqPMinorSEquivalentMaxMin} is true, which in turn implies
\eqref{eqPMinorS}.

In the following, we prove \eqref{eqK1K2Inequality} for any pair
$(k_1,k_2)$ with $k_1\in\cK_1$ and $k_2\in\cK_2$.Let us first
simplify the term $s^{k_1}_i(\mathbf{p})$ by denoting
$s^{k_1}_i(\mathbf{p})=[A_i^{k_1}]^{+}-p^{k_1}_i$, where
$A_i^{k_1}\triangleq{\sigma_i}-\frac{n^{k_1}+\sum_{j\ne
i}|h_j^{k_1}|^2p_j^{k_1}}{|h_i^{k_1}|^2}$. We can verify
that:{\small
\begin{align}
0\le [A_i^{k_1}]^{+}-p_i^{k_1}= A_i^{k_1}-p_i^{k_1}\
,\forall~k_1\in\cK_1.\label{eqK1Inequality}
\end{align}}
Similarly, the term $s^{k_2}_i(\mathbf{p})$ can be simplified as
$s^{k_2}_i(\mathbf{p})=[A_i^{k_2}]^{+}-p_i^{k_2}$. In this case, we
have:{\small
\begin{align}
0>[A_i^{k_2}]^{+}-p_i^{k_2}\ge A_i^{k_2}-p_i^{k_2}
,\forall~k_2\in\cK_2.\label{eqK2Inequality}
\end{align}}
Due to \eqref{eqK1Inequality} and \eqref{eqK2Inequality}, in order
to prove \eqref{eqK1K2Inequality}, it suffices to prove that there
exists $0<M^{(k1,k2)}_i<\infty$ such that:{\small
\begin{align}
&\triangledown_{p_i^{k_1}}P(\mathbf{p})-\triangledown_{p_i^{k_2}}P(\mathbf{p})
\ge M^{(k1,k2)}_i
\left((A_i^{k_1}-p_i^{k_1})-(A_i^{k_2}-p_i^{k_2})\right)\label{eqA1A2Inequality}.
\end{align}}
Notice that from the definition, for any $k\in\cK$ we have $
-\left(A_i^{k}-p_i^{k}\right)+{\sigma_i}=\frac{1}{\triangledown_{p_i^{k}}P(\mathbf{p})}$.
Then the above inequality can be simplified to:{\small
\begin{align}
\triangledown_{p_i^{k_1}}P(\mathbf{p})-\triangledown_{p_i^{k_2}}P(\mathbf{p})
\ge\left(\frac{1}{\triangledown_{p_i^{k_2}}P(\mathbf{p})}-\frac{1}{\triangledown_{p_i^{k_1}}P(\mathbf{p})}\right)M^{(k1,k2)}_i\label{eqB1B2}.
\end{align}}
From \eqref{eqK1Inequality} and \eqref{eqK2Inequality} we have that
{\small $ 0\ge-\left(A_i^{k_1}-p_i^{k_1}\right)$} and {\small
$-\left(A_i^{k_2}-p_i^{k_2}\right)>0$}. Therefore we must have $
\triangledown_{p_i^{k_1}}P(\mathbf{p})>\triangledown_{p_i^{k_2}}P(\mathbf{p})$.
Consequently, \eqref{eqB1B2} is equivalent to {\small $
M^{(k_1,k_2)}_i\le \triangledown_{p_i^{k_1}}P(\mathbf{p})\times
\triangledown_{p_i^{k_2}}P(\mathbf{p}). $} Finding such
$M^{(k_1,k_2)}_i>0$ is always possible, as
$\triangledown_{p_i^{k}}P(\mathbf{p})$ is bounded away from 0 for
any $k$ (cf. \eqref{eqPotentialGradient}, and note
$|h_i^k|^2>0~\forall~i,k$ and $n^k>0~\forall~k$).

Now that we can always find a constant $M^{(k_1,k_2)}_i>0$ that
satisfies \eqref{eqK1K2Inequality}, we can take
$M_i=\min_{k_1,k_2\in\mathcal{K}}M_i^{(k_1,k_2)}$ to ensure
\eqref{eqPMinorS}. Thus, taking $M=\min_{i\in\mathcal{N}}M_i$,
\eqref{eqM} is true, and the first part of the proposition is
proved.

The second part of the proposition is straightforward. Due to space
limit, we do not show the proof here.
\end{proof}

\newtheorem{L2}{Lemma}
\begin{L1}\label{lemmaLip}
{\it For two vectors $\mathbf{p}\in\mathcal{P}$ and
$\bar{\mathbf{p}}\in\mathcal{P}$, there must exist constants
$0<D<\infty$, $0<K<\infty$ such that
\begin{align}
\|\mathbf{s}(\mathbf{p})-\mathbf{s}(\bar{\mathbf{p}})\|&\le
D\|\mathbf{p}-\bar{\mathbf{p}}\|{\textrm{~and~}} \|\triangledown
P(\mathbf{p})-\triangledown P(\bar{\mathbf{p}})\|\le
K\|\mathbf{p}-\bar{\mathbf{p}}\|.\nonumber
\end{align}}
\end{L1}
\begin{proof}
First note that if for all $i$, there exists a constant
$0<D_i<\infty$, such that: $
\|\mathbf{s}_i(\mathbf{p})-\mathbf{s}_i(\bar{\mathbf{p}})\|\le
D_i\|\mathbf{p}-\bar{\mathbf{p}}\|\label{eqSLessThanP}$,  then the
first inequality in the lemma is true for
$D=\sqrt{N}\times\max_{i\in\mathcal{N}}D_i$.
Define a $K\times 1$ vector ${\bf{insr}}_i(\bp_{-i})$ where its
$k$th element is given as  $ \frac{n^k+\sum_{j\ne
i}|h_j^k|^2p_j^k}{|h_i^k|^2}$; let $[\cdot]_{\cP_i}$ denote the
projection to the feasible set $\cP_i$. Then from \cite[Lemma
1]{scutari08b}, we have that
$\mathbf{\Phi}_i(\mathbf{p}_{-i})=\left[-
{\bf{insr}}_i\right]_{\mathcal{P}_i}$. Notice that we have{\small
\begin{align}
&\|{\bf{insr}}_i(\bp_{-i})-{\bf{insr}}_i(\bar{\bp}_{-i})\|\le
\Bigg(\sum_{k=1}^{K}\left|\frac{\sum_{j\ne
i}|h_j^k|^2|p_j^k-\bar{p}^k_j|}{|h_i^k|^2}\right|^2\Bigg)^{1/2}\nonumber\\
&\le\max_k\left\{\frac{\sqrt{\sum_{j\ne
i}|h^k_j|^4}}{|h^k_i|^2}\right\}\Bigg(\sum_{k=1}^{K}(N-1){\sum_{j\ne
i}|p_j^k-\bar{p}^k_j|^2}\Bigg)^{1/2}\nonumber\\
&\le
\bar{D_i}\|\mathbf{p}-\bar{\mathbf{p}}\|\label{eqDifferenceINSR}
\end{align}}
\hspace{-0.1cm}where {\small$\bar{D_i}\triangleq
\max_k\left\{\frac{\sqrt{\sum_{j\ne
i}|h^k_j|^4}}{|h^k_i|^2}\right\}\sqrt{N-1}$}. Then we have
that{\small
\begin{align}
&\|\mathbf{s}_i(\mathbf{p})-\mathbf{s}_i(\bar{\mathbf{p}})\|
=\|\mathbf{\Phi}_i(\mathbf{p}_{-i})-\mathbf{\Phi}_i(\bar{\mathbf{p}}_{-i})+\mathbf{p}_i-\bar{\mathbf{p}}_i\|\nonumber\\
&\stackrel{(a)}\le
\|{\bf{insr}}_i(\bp_{-i})-{\bf{insr}}_i(\bar{\bp}_{-i})\|+
\|\mathbf{p}_i-\bar{\mathbf{p}}_i\|\nonumber\\
&\le \bar{D}_i\|\bp-\bar{\bp}\|+
\|\mathbf{p}_i-\bar{\mathbf{p}}_i\|\le
(\bar{D}_i+1)\|\bp-\bar{\bp}\|
\end{align}}
 where $(a)$ is because of the triangular inequality and the
non-expansiveness of the Euclidean norm. Thus, taking
$D_i=\bar{D}_i+1$, we have that $
\|\mathbf{s}_i(\mathbf{p})-\mathbf{s}_i(\bar{\mathbf{p}})\|\le
D_i\|\mathbf{p}-\bar{\mathbf{p}}\| $.

The second inequality in the Lemma can be shown similarly.
\end{proof}


Using Lemma \ref{lemmaAscend} and Lemma \ref{lemmaLip}, Proposition
\ref{propAIWF} can be shown by slightly generalizing the existing
result \cite[Proposition 3.5]{bertsekas96}, which proves the
convergence for a family of gradient methods with diminishing
stepsizes for {\it unconstrained problems}. The generalization of
this cited result to the current {\it constrained} case is to some
extent straightforward, and we omit the proof due to space
limitations.

\vspace{-0.3cm}
\subsection{Proof of Theorem
\ref{theoremConvergenceJASPA}}\label{appProofJASPA}

We first introduce some notations. Define a vector
$\mathbf{c}^{(t)}\in\cW^N$ such that if $\mathbf{b}_i^{(t)}[w]=1$
then $\mathbf{c}^{(t)}[i]=w$. Define a set $\mathcal{A}$ as:
$\mathcal{A}\triangleq\{\mathbf{a}: \mathbf{a} \textrm{~appears
infinitely often in~} \{\mathbf{a}^{(t)}\}_{t=1}^{\infty}\}$.

{\bf Step 1):} We first show that there must exist a NE association
profile $\mathbf{a}^{*}\in\mathcal{A}$. Let us pick any
$\bta\in\mathcal{A}$. Suppose $\bta$ is not a NE association
profile, and in iteration $T$, $\ba^{(T)}=\bta$.

From step 3) of the JASPA, for a specific MU $i$, all the APs
$w\in\mathcal{W}^{(T+1)}_i$ must satisfy \eqref{eqBetterAP}. 
From step 2) of the algorithm, for AP $\bta[i]$, we have
$\bp_{\bta[i]}^{(T+1)}\in\mathcal{E}_{\bta[i]}(\bta)$. From
Corollary \ref{corollaryPotentialMaximization}, we have that
$\bp_{\bta[i]}^{(T+1)}$ must be a NE for the single AP power
allocation game $G$ with the set of players $\{j:
\ba^{(T)}[j]=\bta[i]\}$. Utilizing the definition of the NE in
\eqref{eqDefineNE}, we must have{\small
\begin{align}
\hspace{-0.5cm}\max_{\mathbf{p}_{i,\bta[i]}\in\mathcal{F}_{i,\bta[i]}}
R_i\bigg(\mathbf{p}_{i,\bta[i]},\mathbf{p}^{(T+1)}_{\bta[i]};\bta[i]\bigg)=
R_i\bigg(\mathbf{p}_{\bta[i]}^{(T+1)};{\bta[i]}\bigg).\nonumber
\vspace{-0.5cm}
\end{align}}
That is, $\bta[i]\in\mathcal{W}^{(T+1)}_i$. From step 3) of the
algorithm, we see that each $w\in\mathcal{W}^{(T+1)}_i$, in
particular $\bta[i]$, has the probability of
$\frac{1}{|\mathcal{W}_i^{(T+1)}|}>\frac{1}{W}$ of being selected as
$\bc^{(T)}[i]$. From step 4)--step 5) of the algorithm, we see that
$\ba^{(T+1)}[i]=\bc^{(T)}[i]$ with probability at least
$\frac{1}{M}$. It follows that{\small
\begin{align}
{\rm Prob}\left(\ba^{(T+1)}[i]=\ba^{T}[i]\right)> \frac{1}{M\times
W}.
\end{align}}
Suppose $\bta$ is not a NE, then there must exist a MU $j$ such that
there is an AP $w_j\in\mathcal{W}^{(T+1)}_j$ that satisfies $w_j\ne
\bta[j]$ and {\small
\begin{align}
\hspace{-0.5cm}\max_{\mathbf{p}_{j,w_j}\in\mathcal{F}_{i,w_j}}
R_j\bigg(\mathbf{p}_{j,w_j},\mathbf{p}^{(T+1)}_{w_j};w_j\bigg)>
R_j\bigg(\mathbf{p}_{\bta[j]}^{(T+1)};{\bta[j]}\bigg).\nonumber
\vspace{-0.5cm}
\end{align}}
Following the same argument as in the previous paragraph, we have
${\rm Prob}\left(\ba^{(T+1)}[j]=w_j\right)>\frac{1}{M\times W}$. It
follows that{\small
\begin{align}
{\rm
Prob}\left(\ba^{(T+1)}=[w_j,\ba^{(T)}_{-j}]\right)>\big(\frac{1}{M\times
W}\big)^N
\end{align}}
The proof of Theorem \ref{theoremExistence} suggests that if a
single MU switches to an AP that increases its rate, then the system
potential increases. In our current context, this says if
$\ba^{(T+1)}=[w_j,\ba^{(T)}_{-j}]$, then
$\bar{P}(\ba^{(T)})<\bar{P}(\ba^{(T+1)})$.

Starting from iteration $T$, with positive probability, in each
iteration a single MU switches to its preferred AP. The potentials
generated in this way is strictly increasing. This process, however,
will stop at a {\it finite} time index $T^*$ such that no MU is
willing to switch. Consequently, $\ba^*=\ba^{T^*}$ is an equilibrium
association profile. The finiteness of $T^*$ is from the finiteness
of distinctive association profiles (and hence the finiteness of
possible values for $\bar{P}(\ba)$). Such finiteness combined with
the fact that each step of of the above process happens with
non-zero probability imply that the probability of reaching $\ba^*$
from $\ba^{(T)}=\bta$ is non-zero.

We conclude that with {\it non-zero} probability, a NE profile
$\mathbf{a}^*$ will appear after $\mathbf{a}^{(T)}$ in {\it finite}
steps. Combined with the assumption that $\mathbf{a}^{(T)}=\bta$
appears infinitely often, we must also have that
$\mathbf{a}^*\in\mathcal{A}$. 

{\bf Step 2):} We can then show that the sequence
$\left\{\mathbf{a}^{(t)}\right\}_{t=1}^{\infty}$ converges to an
equilibrium profile $\mathbf{a}^*$. This step can be shown using the
same argument as in \cite[Theorem 2]{hong11_infocom}. Due to space
limitation, we choose not to reproduce the proof here.

\vspace{-0.3cm}
\subsection{Proof of Proposition
\ref{propIO2}}\label{appProofJJASPASubsequenceConverge}
\begin{proof}
For a $w\in\mathcal{W}$, let $\ctN_w\triangleq\{i:\bta[i]=w\}$. Let
$\{\tilde{t}_n\}_{n=1}^{\infty}$ be a subsequence in which the
subset of MUs $\ctN_w$ is associated with AP $w$. Clearly,
$\{t_n\}_{n=1}^{\infty}$ is a subsequence of
$\{\tilde{t}_n\}_{n=1}^{\infty}$. From the J-JASPA algorithm, we see
that at each $\tilde{t}_n$, \eqref{eqJJASPAPowerUpdate} implements
the single AP A-IWF algorithm with the {\it fixed} set of MUs
$\ctN_w$. Therefor, Proposition \ref{propAIWF} implies that the
subsequence
$\left\{\mathbf{p}^{(\tilde{t}_n)}_w\right\}_{n=1}^{\infty}$
(consequently $\left\{\mathbf{p}^{(t_n)}_w\right\}_{n=0}^{\infty}$)
converges to $\mathbf{p}^*_w$, an optimizer of the problem
$\max_{\bp_w\in\mathcal{F}_w(\bta)}P_w(\bp_w;\bta)$. From Corollary
\ref{corollaryPotentialMaximization} and the fact that
$\ba^{(t_n)}=\bta$ for all $n$, we obtain
$\lim_{n\to\infty}P_w\left(\mathbf{p}_w^{(t_n)};
\ba^{(t_n)}\right)=\bar{P}_w(\bta)$ and
$\lim_{n\to\infty}P\left(\mathbf{p}^{(t_n)},\ba^{(t_n)}\right)=\bar{P}(\bta)$.
\end{proof}

\vspace{-0.3cm} \subsection{Proof of Proposition
\ref{propInclusion}}\label{app3}
\begin{proof}
From Proposition \ref{propIO2}, we have that for a given
$\bta\in\mathcal{A}$,
$\lim_{n\to\infty}\mathbf{p}^{(\tilde{t}_n)}_w=\mathbf{p}_w^*,\forall~w$,
which implies that
$\lim_{n\to\infty}\mathbf{I}_{i,w}^{(\tilde{t}_n)}=\mathbf{I}_{i,w}^*(\bta)$
and
$\lim_{n\to\infty}\bhI_{i,w}^{(\tilde{t}_n)}=\mathbf{I}_{i,w}^*(\bta),
\forall~i$. The latter equality combined with the continuity of the
rate function $R^*_i\left( \widehat{\mathbf{I}}_{i,w};w\right)$ with
respect to $\widehat{\mathbf{I}}_{i,w}$, and the continuity of the
function $R_i\left(\mathbf{p}_w, w\right)$ with respect to
$\mathbf{p}_w$, further implies that, for any $\delta>0$, there must
be a constant $N(\delta)$ such that for all $n
>N(\delta)$, the following are true:{\small
\begin{align}
\left|R^*_i\left(
\widehat{\mathbf{I}}^{(\tilde{t}_n)}_{i,w};w\right)-R^*_i\left(
{\mathbf{I}}^*_{i,w}(\bta); w\right)\right|&<\delta,\nonumber\\
\left|R_i(\bp_{\bta[i]}^{(\tilde{t}_n)};
\bta[i])-{R}_i(\mathbf{p}_{\bta[i]}^*;\bta[i])\right|&<\delta.
\end{align}}
For any $w\ne\mathbf{a}[i]$ satisfying $w\in
\cB_i\left({\mathbf{I}}^*_i(\mathbf{a}),{\mathbf{a}}\right)$, there
must exit a $\epsilon_w>0$ such that:{\small
\begin{align}
R^*_i\left( {\mathbf{I}}^*_{i,w}(\mathbf{a});w\right)-R_i\left(
{\mathbf{I}}^*_{i,\mathbf{a}[i]}(\mathbf{a});\mathbf{a}[i]\right)\ge
{\epsilon}_w.
\end{align}}
Take $\epsilon=\min_{w\in
\cB_i\left({\mathbf{I}}^*_i(\bta),{\bta}[i]\right)}\epsilon_w$,
choose a constant $\widehat{\delta}$ satisfying
$0<2\widehat{\delta}<{\epsilon}$, and let
$N_i^*(\mathbf{a})\triangleq N(\widehat{\delta})$. For simplicity of
notation, write $R^{(t)}_i$ instead of
$R_i\left(\mathbf{p}^{(t)}_{\mathbf{a}^{(t)}[i]},
\mathbf{a}^{(t)}[i]\right)$. We have that for all $n>N_i^*(\bta)$,
the following is true:{\small
\begin{align}
&R^*_i\left( {\mathbf{I}}^*_{i,w}(\bta);w\right)-R_i\left(
{\mathbf{I}}^*_{i,\bta[i]}(\bta);\bta[i]\right)\nonumber\\
&= R_i\left( {\mathbf{I}}^*_{i,w}(\bta);w\right) + R^*_i\left(
\widehat{\mathbf{I}}^{(\tilde{t}_n)}_{i,w};w\right)-R^*_i\left(
\widehat{\mathbf{I}}^{(\tilde{t}_n)}_{i,w};w\right) \nonumber\\
&\quad + \widehat{R}^{(\tilde{t}_n)}_i-\widehat{R}^{(\tilde{t}_n)}_i
 - R_i\left(
{\mathbf{I}}^*_{i,\bta[i]}(\bta);\bta[i]\right)\nonumber\\
&\le R^*_i\left(
\widehat{\mathbf{I}}^{(\tilde{t}_n)}_{i,w};w\right)-\widehat{R}^{(\tilde{t}_n)}_i+
\left|R^*_i\left( {\mathbf{I}}^*_{i,w}(\bta);w\right)-R^*_i\left(
\widehat{\mathbf{I}}^{(\tilde{t}_n)}_{i,w};w\right)
\right|\nonumber\\
&\quad+\left|\widehat{R}^{(\tilde{t}_n)}_i- R_i\left(
{\mathbf{I}}^*_{i,\bta[i]}(\bta);\bta[i]\right)\right|\nonumber\\
&\le R^*_i\left(
\widehat{\mathbf{I}}^{(\tilde{t}_n)}_{i,w};w\right)-\widehat{R}^{(\tilde{t}_n)}_i+\widehat{\delta}+\widehat{\delta}.
\end{align}}
Consequently, we have that for all $n>N_i^*(\bta)$, $
R^*_i\left(\widehat{\mathbf{I}}^{(\tilde{t}_n)}_{i,w};w\right)-\widehat{R}^{(\tilde{t}_n)}_i
\ge {\epsilon}-2\widehat{\delta}>0, $ which implies that $w$ must be
in the set
$\cB_i\left(\widehat{\mathbf{I}}^{(\widetilde{t}_n)}_i,\widehat{\mathbf{a}}^{(\widetilde{t}_n)}[i]\right)$.
The claim is proved.
\end{proof}

\vspace{-0.3cm}
\subsection{Proof of Theorem
\ref{theoremConvergenceJJASPA}}\label{appProofJJASPA}
\begin{proof}
Consider the sequence $\{\left(\mathbf{a}^{(t)},
\mathbf{p}^{(t)}\right)\}_{t=1}^{\infty}$. Choose
$\widetilde{\mathbf{a}}$ to be any system association profile that
satisfies: $
\widetilde{\mathbf{a}}\in\arg\max_{\mathbf{a}\in\mathcal{A}}\bar{P}(\mathbf{a}).
$ Again let $\left\{\tilde{t}_n: \ba^{(\tilde{t}_n)}=\bta
~\textrm{and}~ \widehat{\mathbf{a}}^{(\tilde{t}_n)}=\bta\right\}$.
From Proposition \ref{propIO2} we have that the sequence
$\bp^{(\tilde{t}_n)}$ converges, i.e.,
$\lim_{n\to\infty}\bp^{(\tilde{t}_n)}=\bp^*$. We first show that
$\left(\widetilde{\mathbf{a}},\mathbf{p}^*\right)$ is a JEP.

Suppose $\left(\widetilde{\mathbf{a}},\mathbf{p}^*\right)$ is not a
JEP, then there exists a MU $\check{i}$, and a
$\check{w}\ne\widetilde{\mathbf{a}}[{\check{i}}]$ such that
$\check{w}\in
\cB_{\check{i}}\left({\mathbf{I}}^*_{\check{i}}(\widetilde{\mathbf{a}}),{\widetilde{\mathbf{a}}}[i]\right)$.
This implies that there exists an $\underline{\epsilon}>0$ such
that:{\small
\begin{align}
R^*_{\check{i}}\left(
{\mathbf{I}}^*_{{\check{i}},\check{w}}(\widetilde{\mathbf{a}});\check{w}\right)-R_{\check{i}}\left(
{\mathbf{I}}^*_{\check{i},\widetilde{\mathbf{a}}[{\check{i}}]}(\widetilde{\mathbf{a}});
\widetilde{\mathbf{a}}[{\check{i}}]\right)\ge \underline{\epsilon}.
\end{align}
} Define a new association profile
$\check{\mathbf{a}}\triangleq[\check{w},\mathbf{a}_{-\check{i}}]$.
Following the similar steps as in the proof of Theorem
\ref{theoremExistence}, we can show that: $
\bar{P}(\widetilde{\mathbf{a}})<\bar{P}(\check{\mathbf{a}})$. It is
clear that if $\check{\mathbf{a}}\in\mathcal{A}$, then the previous
inequality is a contradiction to the assumption that $
\widetilde{\mathbf{a}}\in\arg\max_{\mathbf{a}\in\mathcal{A}}\{\bar{P}(\mathbf{a})\}
$. In the following, we show that
$\check{\mathbf{a}}\in\mathcal{A}$, which completes the proof.

From Proposition \ref{propInclusion}, there exists a
$N_{\check{i}}^*(\widetilde{\mathbf{a}})$ large enough that for all
${n}>N_{\check{i}}^*(\widetilde{\mathbf{a}})$, $\check{w}\in
\cB_{\check{i}}\left(\widehat{\mathbf{I}}^{(\tilde{t}_n)}_{\check{i}},
\widehat{\mathbf{a}}^{(\tilde{t}_n)}[{\check{i}}]\right)$. Take any
$n>N_{\check{i}}^*(\widetilde{\mathbf{a}})$. From the definition, in
iteration $\tilde{t}_n$, $\bta$ is sampled, that is,
$\widehat{\mathbf{a}}^{(\tilde{t}_n)}=\widetilde{\mathbf{a}}$. From
step 5) of the J-JASPA algorithm, we see that with non-zero
probability, in iteration $\tilde{t}_n+1$, all MUs $j\ne
{\check{i}}$ stay in $\widetilde{\mathbf{a}}[j]$, and MU
${\check{i}}$ chooses to switch to $\check{w}$. This implies that
the association profile $\check{\mathbf{a}}$ happens with non-zero
probability in every time instance $\tilde{t}_n+1$. Because
$\{\tilde{t}_n\}$ is an infinite sequence, $\check{\mathbf{a}}$
happens infinitely often, i.e., $\check{\mathbf{a}}\in\mathcal{A}$.

In summary, we conclude that $\widetilde{\mathbf{a}}$ must be a NE
association profile, and thus,
$\left(\widetilde{\mathbf{a}},\mathbf{p}^*(\widetilde{\mathbf{a}})\right)$
is a JEP.

Finally, following the proofs of Theorem
\ref{theoremConvergenceJASPA}, we can show similarly that the
sequence $\left\{(\mathbf{a}^{(t)},
\mathbf{p}(\mathbf{a}^{(t)}))\right\}_{t=1}^{\infty}$ generated by
the J-JASPA converges to a JEP with probability 1.
\end{proof}

\vspace{-0.2cm} {\small
\bibliographystyle{IEEEbib}
\bibliography{ref}
}
\end{document}


\title{Proof of Proposition 1}

\maketitle

\section{Proof of Proposition 1}\label{appPropAIWF}
\emph{Proposition 1: {\it If all the MUs in the network employ A-IWF
algorithm, then starting from any feasible initial power allocations
$\mathbf{p}^{(0)}\in\mathcal{P}$, their individual power profiles
converge to the set of NE of game $G$}}

{\it Proof}: To conform to the convention in convex optimization, we
define the function $F(\mathbf{p})=-P(\mathbf{p})$, and we see that
$F(\mathbf{p})$ is convex.

Then from the well known Descent Lemma (Lemma 2.1 in
\cite{bertsekas97}), and Lemma 2 we have that:
\begin{align}
\hspace{-0.3cm}F(\mathbf{p}^{(t+1)})&\le
F(\mathbf{p}^{(t)})+\alpha^{(t)}\mathbf{s}(\mathbf{p}^{(t)})^{\intercal}\triangledown
F(\mathbf{p}^{(t)})+\frac{K}{2}(\alpha^{(t)})^2\|\mathbf{s}(\mathbf{p}^{(t)})\|^2\label{eqTylor}\\
&\le
F(\mathbf{p}^{(t)})-\alpha^{(t)} M\|\mathbf{s}(\mathbf{p}^{(t)})\|^2+\frac{K}{2}(\alpha^{(t)})^2
\|\mathbf{s}(\mathbf{p}^{(t)})\|^2\nonumber\\
&=F(\mathbf{p}^{(t)})-\alpha^{(t)}
\|\mathbf{s}(\mathbf{p}^{(t)})\|^2(M-\alpha^{(t)} \frac{K}{2}).
\end{align}
From the assumption that $\alpha^{(t)}$ goes to $0$, there must
exist a $t_0$ such that for all $t\ge t_0$, we have
$M-\alpha^{(t)}\frac{K}{2}>0$, and $F(\mathbf{p}^{(t)})$ is
monotonically decreasing. Combined with the fact that
$F(\mathbf{p}^{(t)})$ is lower bounded, then
$\{F(\mathbf{p}^{(t)})\}_{t=1}^{\infty}$ is a convergent sequence.

From \eqref{eqTylor}, we have that
\begin{align}
F(\mathbf{p}^{(T+1)})&\le
F(\mathbf{p}^{(0)})+\sum_{t=0}^{T}\alpha^{(t)}\mathbf{s}(\mathbf{p}^{(t)})^{\intercal}\triangledown
F(\mathbf{p}^{(t)})+\sum_{t=0}^{T}\frac{K}{2}(\alpha^{(t)})^2\|\mathbf{s}(\mathbf{p}^{(t)})\|^2.\nonumber
\end{align}
It is clear that  $\|\mathbf{s}(\mathbf{p}^{(t)})\|^2$ is upper
bounded, and we have
$\lim_{T\to\infty}\sum_{t=1}^{T}(\alpha^{(t)})^2<\infty$, so we must
have
$\lim_{T\to\infty}\sum_{t=0}^{T}\frac{K}{2}(\alpha^{(t)})^2\|\mathbf{s}(\mathbf{p}^{(t)})\|^2<\infty$.
Because $\lim_{T\to\infty}F(\mathbf{p}^{(T+1)})$ converges, we must
have
\begin{align}
\lim_{T\to\infty}\sum_{t=0}^{T}\alpha^{(t)}\mathbf{s}(\mathbf{p}^{(t)})^{\intercal}\triangledown
F(\mathbf{p}^{(t)})&>-\infty, {\rm~or~equivalently,~}
\lim_{T\to\infty}\sum_{t=0}^{T}\alpha^{(t)}\mathbf{s}(\mathbf{p}^{(t)})^{\intercal}\triangledown
P(\mathbf{p}^{(t)})&<\infty.\label{eqSummability}
\end{align}
Combining with Lemma 1, we have
\begin{align}
M\lim_{T\to\infty}\sum_{t=0}^{T}\alpha^{(t)}\|\mathbf{s}(\mathbf{p}^{(t)})\|^2
\le\lim_{T\to\infty}\sum_{t=0}^{T}\alpha^{(t)}\mathbf{s}(\mathbf{p}^{(t)})^{\intercal}\triangledown
P(\mathbf{p}^{(t)})<\infty.\nonumber
\end{align}
Using the fact that the sequence $\{\alpha^{(t)}\}$ is not summable,
we must have
$\lim\inf_{t\to\infty}\|\mathbf{s}(\mathbf{p}^{(t)})\|=0$. We show
in the following that in fact we have a stronger result that
$\lim_{t\to\infty}\|\mathbf{s}(\mathbf{p}^{(t)})\|=0$. Suppose not,
then $\lim\sup_{t\to\infty}\|\mathbf{s}(\mathbf{p}^{(t)})\|>0$. In
this case there must exist a $\epsilon>0$ such that the subsequences
$\{\tau_n: \|\mathbf{s}(\mathbf{p}^{(\tau_n)})\|<\epsilon,
||\mathbf{s}(\mathbf{p}^{(\tau_n+1)})||\ge\epsilon\}$ and $\{u_n:
\epsilon\le \|\mathbf{s}(\mathbf{p}^{(t)})\|\le 2\epsilon,
\forall~t\in(\tau_n, u_n-1),~ \|\mathbf{s}(\mathbf{p}^{(u_n)})\|>
2\epsilon\}$ are both infinite.

For a specific $n$, Lemma 2 implies that the following is true:
\begin{align}
\|\mathbf{s}(\mathbf{p}^{(\tau_n+1)})\|-\|\mathbf{s}(\mathbf{p}^{(\tau_n)})\|&\le
D\|\mathbf{p}^{(\tau_n+1)}-\mathbf{p}^{(\tau_n)}\|\nonumber\\
&\le D\alpha^{(\tau_n)}\|\mathbf{s}(\mathbf{p}^{(\tau_n)})\|.
\end{align}
Thus, there exists a constant $N^*$ such that for all $n>N^*$, we
must have
$\|\mathbf{s}(\mathbf{p}^{(\tau_n)})\|\ge\frac{\epsilon}{2}$.

We also have the following series of inequalities:
\begin{align}
\epsilon&<
\|\mathbf{s}(\mathbf{p}^{(u_n)})\|-\|\mathbf{s}(\mathbf{p}^{(\tau_n)})\|
\le D\|\mathbf{p}^{(u_n)}-\mathbf{p}^{(\tau_n)}\|\nonumber\\
&\le
D\sum_{t=\tau_n}^{t=u_n-1}\alpha^{(t)}||\mathbf{s}(\mathbf{p}^{(t)})||\le
D\sum_{t=\tau_n}^{t=u_n-1}\alpha^{(t)}2\epsilon
\end{align}
which implies
\begin{align}
\frac{1}{2D}<
\sum_{t=\tau_n}^{t=u_n-1}\alpha_t.\label{eqContradictionD}
\end{align}
From our previous derivation, we also have $
\lim_{T\to\infty}\sum_{t=0}^{T}\alpha^{(t)}\|\mathbf{s}(\mathbf{p}^{(t)})\|^2<\infty.
$ Then for any $\delta>0$ there must exists a $\widehat{N}(\delta)$
such that for all $n>\widehat{N}(\delta)$: $
\sum_{t=\tau_n}^{t=u_n-1}\alpha^{(t)}\|\mathbf{s}(\mathbf{p}^{(t)})\|^2\le
\delta. $

Take $\delta=\frac{\epsilon^2 }{8 D}$, and take $n>\max\left\{{N^*,
\widehat{N}(\frac{\epsilon^2 }{8 D})}\right\}$, then we have
\begin{align}
\frac{\epsilon^2}{4}\sum_{t=\tau_n}^{t=u_n-1}\alpha^{(t)}
\le\sum_{t=\tau_n}^{t=u_n-1}\alpha^{(t)}\|\mathbf{s}(\mathbf{p}^{(t)})\|^2
\le\frac{\epsilon^2 }{8 D}
\end{align}
which implies $\sum_{t=\tau_n}^{t=u_n-1}\alpha^{(t)}\le \frac{1}{2
D}$. This is a contradiction to \eqref{eqContradictionD}. Thus, we
conclude that $\lim\sup_{t\to\infty}\|\mathbf{s}(\mathbf{p}^{(t)})\|=0$,
and consequently
$\lim_{t\to\infty}\|\mathbf{s}(\mathbf{p}^{(t)})\|=0$.

From $\lim_{t\to\infty}\|\mathbf{s}(\mathbf{p}^{(t)})\|=0$ we see
that the limit point $\mathbf{p}^*$ of any converging subsequence of
$\{\mathbf{p}^{(t)}\}$, must satisfy
$\bfPhi(\mathbf{p}^*)=\mathbf{p}^*$, which ensures that
$\mathbf{p}^*$ is a NE of the game $G$. Consequently $\mathbf{p}^*$
must maximize the function $P(\mathbf{p})$ (from Corollary 1), and
this implies that the entire sequence
$\{P(\mathbf{p}^{(t)})\}_{t=1}^{\infty}$ converges to the value
$\max_{\mathbf{p}\in\mathcal{P}}P(\mathbf{p})$.

\bibliographystyle{IEEEbib}
\bibliography{ref}